\documentclass[12pt,preprint]{aastex}

\newcommand{\gcc}{\ensuremath{\mathrm{g\ cm^{-3} }}}
\newcommand{\cc}{\ensuremath{\mathrm{cm^{-3} }}}

\newcommand{\cm}{\ensuremath{\mathrm{cm}}}

\newcommand{\cmss}{\ensuremath{\mathrm{cm \ s^{-2}}}}
\newcommand{\kpc}{\ensuremath{\mathrm{kpc}}}
\newcommand{\Myr}{\ensuremath{\mathrm{Myr}}}
\newcommand{\keV}{\ensuremath{\mathrm{keV}}}

\newcommand{\FLASH}{{\sc{Flash}}{}}

\usepackage{epsf,color}

\epsfverbosetrue

\begin{document} 

\title{Morphology of Rising Hydrodynamic and Magneto-hydrodynamic Bubbles
       from Numerical Simulations}

\shorttitle{Simulations of Rising Gas Bubbles}

\author{
    K.~Robinson\altaffilmark{1,3},
    L.~J.~Dursi\altaffilmark{2,3},
    P.~M.~Ricker\altaffilmark{4,5},
    R.~Rosner\altaffilmark{2,3,6},
    A.~C.~Calder\altaffilmark{2,3},
    M.~Zingale\altaffilmark{7},
    J.~W.~Truran\altaffilmark{2,3},
    T.~Linde\altaffilmark{2,3}, 
    A.~Caceres\altaffilmark{3,6},
    B.~Fryxell\altaffilmark{2,3},
    K.~Olson\altaffilmark{3,8},
    K.~Riley\altaffilmark{3},
    A.~Siegel\altaffilmark{3},
    N.~Vladimirova\altaffilmark{3}}

\altaffiltext{1}{Dept.\ of Physics,
                 Lawrence University, 
                 Appleton, WI 54912}

\altaffiltext{2}{Dept.\ of Astronomy \& Astrophysics, 
                 The University of Chicago, 
                 Chicago, IL  60637}

\altaffiltext{3}{Center for Astrophysical Thermonuclear Flashes, 
                 The University of Chicago, 
                 Chicago, IL  60637}

\altaffiltext{4}{Dept.\ of Astronomy,
                 University of Illinois at Urbana-Champaign,
                 Urbana, IL 61801}

\altaffiltext{5}{National Center for Supercomputing Applications,
                 Urbana, IL 61801}

\altaffiltext{6}{Dept.\ of Physics,
                 The University of Chicago,
                 Chicago, IL 60637}

\altaffiltext{7}{Dept.\ of Astronomy \& Astrophysics,
                 The University of California, Santa Cruz,
                 Santa Cruz, CA 95064}

\altaffiltext{8}{UMBC/GEST Center, NASA/GSFC, 
                 Greenbelt, MD 20771}

\begin{abstract}
Recent {\it Chandra} and {\it XMM-Newton}
observations of galaxy cluster cooling flows have revealed X-ray emission
voids of up to 30~kpc in size that have been identified with 
buoyant, magnetized bubbles.
Motivated by these observations, we have investigated the 
behavior of rising bubbles in stratified atmospheres using the \FLASH{}
adaptive-mesh simulation code.  We present results from 2-D simulations
with and without the effects of magnetic fields, and with varying
bubble sizes and background stratifications.  We find purely hydrodynamic bubbles
to be unstable; a dynamically important magnetic field is required to
maintain a bubble's integrity.  This suggests that, even absent thermal
conduction, for bubbles to be persistent enough to be regularly
observed, they must be supported in large part by magnetic fields.
Thermal conduction unmitigated by magnetic fields can dissipate the bubbles
even faster.   We also observe that the bubbles leave a tail as they rise;
the structure of these tails can indicate the history of the dynamics of
the rising bubble.
\end{abstract}


\section{INTRODUCTION}
\label{sec:introduction}
Cooling flows have been known for some time to exist at the centers of
some clusters of galaxies (see \citealt{fabian1994} for a review).  The
X-ray emissivity of the intracluster medium (ICM) increases with the
square of the plasma density, and as the ICM is very centrally
concentrated, radiative energy losses are greatest at the center of a
cluster. In rich clusters these losses over the life of the cluster can
be significant enough, in the absence of thermal conduction, to cause
the plasma to contract gradually within the cluster potential and flow
inward with velocities of order a few tens of km~s$^{-1}$.  It has been
understood that --- again, in the absence of conduction -- the cooling
gas must become thermally unstable to the formation of a multiphase
medium below about 0.1~\keV, but the fate of this cold gas -- as stars,
cold X-ray-absorbing clouds, or even brown dwarfs -- has been a
mystery. The possibility that some energetic feedback mechanism could
shut off the cooling flows has been subject to debate, as well as the
role played by thermal conduction and magnetic fields.

Recent high-resolution X-ray observations by the {\it Chandra}
and {\it XMM-Newton} satellites have dramatically changed this picture
\citep{boehringeretal2002}.
First, no evidence from spatially resolved spectroscopy has been observed
for the presence of gas colder than about 1--2~\keV{} 
(e.g., \citealt{schmidt2001}; \citealt{petersonetal2001}; \citealt{kaastra2001};
\citealt{tamuraetal2001}; \citealt{matsushita2002}).
Second, high-resolution imaging has provided evidence for large-scale
motions that can heat the ICM (e.g., \citealt{mcnamara2000};
\citealt{mcnamara2001}; \citealt{forman2002}; \citealt{blantonetal2003}).
For years it has been known that many of the
same clusters that harbor cooling flows also contain large
central galaxies with active nuclei. However, the extent to
which these active galactic nuclei (AGN) influence the dynamical
state of the cooling ICM has only become apparent through
the new X-ray observations. These observations show that AGN
produce massive outflows of magnetized plasma that displace
the cooling gas. The `bubbles' thus produced are known to be magnetized
because radio observations show regions of synchrotron
emission coinciding with the regions of low X-ray emission,
and the polarization of this radiation
shows Faraday rotation effects consistent with dynamically
important magnetic fields (e.g., \citealt{allenetal2001};
\citealt{nulsenetal2002}).
The emerging picture is that bubbles represent the late stages of
propagation of the magnetized, relativistic jets produced by AGN
into the ICM, after they have slowed and reached approximate
pressure equilibrium with the ICM \citep{reynoldsetal2002}.

The influence of these magnetized bubbles on the cooling ICM
is still poorly understood. How efficiently does the bubble plasma
mix with the ICM? Does it heat the cooling flow sufficiently to
avoid the formation of multiphase gas and the possible formation
of stars? If so, how -- by radiative heating, magnetic reconnection,
or some other process? What fraction of the total energy budget
of a cluster is contributed by magnetic fields and cosmic rays?
Could these active regions be sites for the acceleration of the
ultra-high-energy cosmic rays seen in some air-shower experiments?

Owing to the geometrical and physical complexity of the AGN-cooling
flow environment, numerical simulations are the best theoretical tools
available for addressing these questions.  While bubbles in liquids are
well-studied (see, e.g., \citealt{recentreview,bubblesreview}), bubbles
in cluster cooling flows should differ from bubbles in liquids in
several important ways.  In particular, molecular forces present in
liquids (e.g., surface tension, viscosity) will not play a significant
role, and magnetic fields may be important.  These differences will
affect the dynamics and stability properties of bubbles.

Several numerical studies have been performed recently
of rising bubbles in the context of cluster cooling flows.
\cite{churazovetal2001} used 2-D hydrodynamic simulations in a cooling-flow
model atmosphere to gain insight into the bubbles observed in M87.
\cite{bruggenkaiser2002,bruggenkaiser2001} 
studied 2-D spherical and elliptical
magnetohydrodynamic (MHD) bubbles rising in a hydrostatic background
medium with an isothermal $\beta$-model density profile
\citep{betamodel},
\begin{equation}
\rho(r) \propto [1 + (r/r_c)^2]^{-3\beta/2}\ ,
\end{equation}
where $r_c \sim 200$~\keV{} is the core radius and $\beta$ was taken to be
0.5.
\cite{bruggenetal2002} performed 3-D hydrodynamic bubble simulations with
continuous energy injection.
More recently, \cite{bassonalexander2003} have performed 3-D hydrodynamic
simulations of both the active and inactive phases of AGN jets propagating
into $\beta$-model density profiles, including radiative cooling.
In addition, several papers have addressed
the detectability in radio and X-rays of cluster bubbles,
including \cite{churazovetal2001}, \cite{bruggenkaiser2001},
\cite{ensslinheinz}, and \cite{sokerblantonsarazin}.
Finally, bubbles in stratified atmospheres have been studied in the
context of contact binary stars by \cite{brandenburg2001}.

In this paper, we study the general question of the behavior of a buoyant
gas bubble rising in a denser gas, investigating the relative
importance of the effects of geometry, stratification, density
contrast, and thermal and magnetic pressure on the stability of the
bubble.  Most of the previous studies (both 2-D and 3-D) have used
finite-difference methods to evolve the gas on relatively
low-resolution grids, up to about 400 zones on a side. 
\cite{newbruggen} and \cite{bruggenkaiser2002} used the \FLASH{} code
\citep{flashcode}, which has an adaptive mesh \citep{paramesh} to
achieve an effective resolution of 4,000$\times$2,000 zones for purely
hydrodynamic calculations.  We also employ \FLASH, but in addition to
hydrodynamic bubbles we study the effects of magnetic fields with
different spatial configurations, exploiting a new MHD module we have
developed for \FLASH. In addition, because we do not treat the bubbles
as self-gravitating, we are able to use for the hydrodynamic case a
version of the piecewise parabolic method (PPM) that has been modified
to handle nearly hydrostatic flows \citep{hsepaper}.  Our initial model
is deliberately simplified as a first step in a research program to
better understand the dynamics of bubbles in clusters; thus we postpone
detailed observational comparisons to a later paper.

The paper is organized as follows: in \S~\ref{sec:initialmodels} we
describe the construction of our initial models and the numerical
methods used. In \S~\ref{sec:hydrodynamicbubbles} we discuss the
results of our purely hydrodynamic calculations. In
\S~\ref{sec:magneticfields} we discuss results of simulations of hot
bubbles rising in background magnetic fields, and in
\S~\ref{sec:magneticbubbles} we consider magnetized bubbles initially
supported partly by force-free magnetic fields.  In
\S~\ref{sec:conclusions} we summarize our conclusions.

\section{INITIAL MODELS}
\label{sec:initialmodels}
We consider as initial models an under-dense circular bubble in an
isothermal, stratified atmosphere with an imposed constant
gravitational acceleration in the $y$-direction.  An ideal equation of
state is used, with a ratio of specific heats $\gamma = 5/3$.  The
sound speed was chosen, through the temperature and mean molecular
weight, so that our standard simulation domain would span approximately
three pressure scale heights.  For most of our simulations, we simulate
2-D boxes in planar geometry.  We vary the stratification by varying the
gravitational acceleration.  We also vary the size of the bubble.

We perform simulations in which the density contrast between the bubble and
the background is caused by a higher temperature in the bubble in
horizontal pressure equilibrium with its surroundings (the `hot bubble'
case) and also in which the pressure support for the under-dense bubble
comes from an azimuthal magnetic field (the `magnetized bubble' case).
In the `hot bubble'  case, we perform simulations with and without a
background magnetic field.   The bubble is in horizontal pressure
equilibrium at every point inside the bubble; since the bubbles
considered here are significant fractions of a pressure scale height,
the bubble's pressure structure is itself somewhat stratified.

The physics in the simulations presented is all scale-free.  For
numerical reasons, we choose scales for the simulations so that numbers
simulated are all near unity.   Thus, our simulations were performed with a
domain size of $40 \times 50$~cm, and the bubble radius of $3$~cm.  The
base density and pressure are taken to be of order unity ($\rho_0
= 1$, $p_0 \approx .64$) and gravitational acceleration is
$0.04~\cmss$.   For comparison to the astrophysical systems of
interest, however, we rescale the results when plotting figures.   We
rescale by scaling the length scales by a factor $L$, the gravitational
acceleration by $G$, and the density by $R$.  In that case, the
hydrodynamic quantities scale as
\begin{eqnarray}
x & = & L x' \\
t & = & \sqrt{\frac{L}{G}} t' \\
v & = & \sqrt{L G} v' \\
g & = & G g' \\
\rho & = & R \rho' \\
p & = & \left ( R G L \right ) p',
\end{eqnarray}
where the primed quantities represent the simulated values, and
unprimed represent the scaled values.   Here, $x$ is a length, $t$
time, $v$ velocity, $g$ the gravitational acceleration, $\rho$ density,
and $p$ pressure.  The values presented here are all scaled by factors
to be roughly consistent with the model of Abel 2052 used in
\cite{sokerblantonsarazin}:  $L = (3 \kpc)/(1~\cm)$, $G = (7 \times
10^{-9}~\cmss)/(.04~\cmss)$, and $R = (4 \times
10^{-26}~\gcc)/(1~\gcc)$.   With these scalings, the physical background
temperature is approximately $1.6 \times 10^7$~K, and the sound
speed is approximately $0.41 \kpc~\Myr^{-1}$.

We perform our simulations with the \FLASH{} code \citep{flashcode}.
The \FLASH{}  code is an adaptive-mesh reactive hydrodynamics code that
has both hydrodynamic and magneto-hydrodynamic solvers.  The \FLASH{}
code's main hydrodynamic solver has undergone a rigorous validation and
verification process \citep{flashcodeVV}.  The method we use for performing
our magnetic calculations is described in \cite{mhdcodepaper}.   The 
two solvers use similar methods, although the MHD solver uses lower-order
reconstructions than the PPM solver, meaning for the same number of grid
points, it is `lower resolution'.  We use the techniques described in
\cite{hsepaper} to maintain hydrostatic equilibrium; in particular, we
use the initialization procedure described therein and the boundary
conditions.  The modification to PPM to provide more accurate stable
solutions was not found to significantly effect the very dynamic
solutions described here, and was not used with either solver.   Unless
otherwise noted, the simulations described here were performed with an
effective grid size of $512 \times 640$ --- with an unrefined grid of
$128 \times 160$ and two further levels of mesh refinement applied
where there are large second derivatives of density or pressure, as
described in \cite{flashcode}.

We use reflecting boundaries on the left and right, and the hydrostatic
boundaries described in \cite{hsepaper} with `outflow' velocities.   In
simulations with magnetic fields, the magnetic boundary conditions are
reflecting on the left and right, and zero-gradient at the top and
bottom.

Because the gas is miscible, there is no surface tension.   We can
estimate the magnitudes of other unmodelled effects such as viscosity,
conduction, and cooling effects \citep{spitzer}.   If the gas is a 
completely ionized hydrogen gas, scaling by the typical values relevant
to this problem allows us to write
\begin{equation}
{\mathrm{Re}} = U L / \nu = 400 \, T_7^{-5/2} \left ( \frac{\ln \Lambda}{32} \right ) 
\left (\frac{n_i}{10^{-2}\cc} \right ) \left ( \frac{U}{0.1~\kpc~\Myr^{-1}} \right )
\left ( \frac{L}{18~\kpc} \right ),
\end{equation}
where the Reynolds number, Re, expresses the relative importance of
inertial to viscous forces, $T_7$ is the temperature in units of $10^7
K$, $\ln \Lambda$ is the Coulomb logarithm, measuring the range of length
scales over which collisions are important, $n_i$ is the number density
of ions, and $\nu$ is the kinematic viscosity.   The length scale used,
$L = 18~\kpc$, is a typical size of the bubbles we study, and the
velocity scale, $U = 0.1~\kpc~\Myr^{-1}$ is, as we will see, a typical
bubble velocity.

Considerable effort has gone into investigating the relatively low
numerical dissipation in PPM and other shock-capturing schemes as well
as the issue of convergence of solutions of these
methods~\citep{sytine00,garnier99,porter94}.  While there is not
complete agreement about these results and their interpretation, it is
apparent that the intrinsic viscosity of methods such as PPM, even at
the highest attainable resolutions, prevents simulation of high
Reynolds number flows.  Because the Reynolds number of the flow is
certainly larger than can be obtained in these simulations (the 18 kpc
bubble is resolved with only 77 points), the omission of an explicit
viscosity is appropriate.

Similarly, one can compute a Peclet number, Pe, which expresses
the relative importance of heat transport by convection to that
of thermal diffusion:
\begin{equation}
{\mathrm{Pe}} = U L / D_{\mathrm{th}} = 9  \, T_7^{-5/2} \left ( \frac{\ln \Lambda}{32} \right ) 
\left (\frac{n_i}{10^{-2}\cc} \right )
\left (\frac{U}{0.1~\kpc~\Myr^{-1}} \right )
\left (\frac{L}{18~\kpc} \right ).
\end{equation}
Here $D_{\mathrm{th}}$ is the thermal diffusivity, which is approximately
\begin{equation}
D_{\mathrm{th}} = 0.209 \, T_7^{5/2} \left ( \frac{\ln \Lambda}{32} \right)^{-1} \left (\frac{n_i}{10^{-2}\cc}\right )^{-1} \kpc^2/\Myr.
\end{equation}
Here we see that thermal diffusion may be significant; it will
certainly be significant at
the interface between the bubble and the ambient medium.   We do not
include thermal conductivity in the simulations, and we discuss heat
diffusion out of the bubble in \S\ref{sec:conclusions}.

Another neglected effect is radiative cooling of any hot bubbles.
If the radiative cooling is dominated by free-free emission, then
we will have
\begin{equation}
\frac{d T}{dt} = 2300 \, T_7^{1/2} \left(\frac{\bar{g}_{\mathrm{ff}}}{1.103}\right) \left (\frac{n_i}{10^{-2}\cc} \right ) K \Myr^{-1},
\end{equation}
where $dT/dt$ is the change in temperature per unit time.  In this
case, gas at 1~\keV~$\approx 10^7 K$ will take time scales of order
$5000 \Myr$ to radiate away significant amounts of its internal energy,
meaning that such radiative cooling can be neglected for the current
simulations.  However, using more realistic cooling rates such
as in \cite{coolingrates}, we have
\begin{equation}
\frac{d T}{dt} = 48000 \, T_7^{-1/2} \left (\frac{n_i}{10^{-2}\cc} \right ) K \Myr^{-1},
\end{equation}
so that the cooling time scale at $10^7~K$ is 210~\Myr, meaning that
cooling would be starting to play a role during these simulations
were it included.

\section{HYDRODYNAMIC BUBBLES}
\label{sec:hydrodynamicbubbles}
\begin{figure}
\plotone{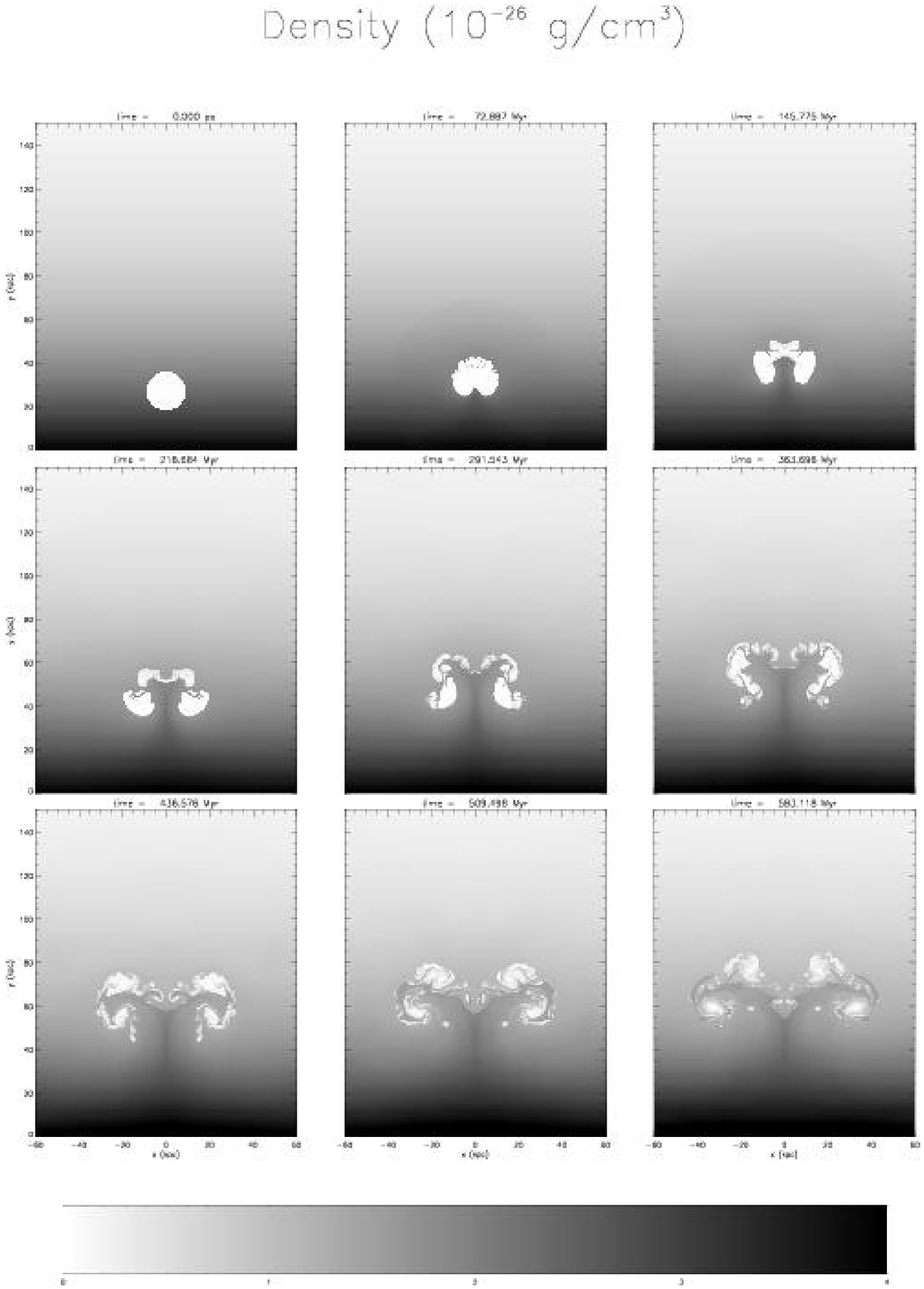}
\caption{The time evolution of our `standard' bubble case - $g = -7\times10^{-9} \cmss$,
density contrast of 100:1, peak density of $\rho = 4 \times 10^{-26} \gcc$,
3.1 pressure scale heights in the 2-D planar domain, and bubble radius of $9~\kpc$.
Plotted is density.}
\label{fig:basecase}
\end{figure}

We first examine our reference case -- a simulation with $g = -7\times
10^{-9}~\cmss$ and a density at the bottom of the simulation domain of
$\rho = 4 \times 10^{-26}~\gcc$.  With these parameters, our box
(150~\kpc\ tall) contains $3.1$ pressure scale heights.   The density
contrast between the bubble and the maximum density in the box is
100:1, and there are no magnetic fields.  The bubble radius is $r =
9~\kpc$, making the whole bubble a little more than one-third of a
pressure scale height in size.  This simulation is in 2D planar
coordinates.  The evolution of density is shown in
Fig.~\ref{fig:basecase}.

As we have scaled our results to Abell 2052, our bubble size (in units
of pressure scale heights) is consistent with that of bubbles observed
in that cluster (approximately 0.35).   In other clusters, bubbles are
seen with sizes of approximately 0.1--0.9 pressure scale heights:
examples are 0.08 in Abell~478\citep{sunjones}; 0.25--0.37 in
Abell~2597\citep{mcnamara2001}; 0.64--0.9 in
Hydra~A\citep{mcnamara2000}.

\begin{figure}
\plotone{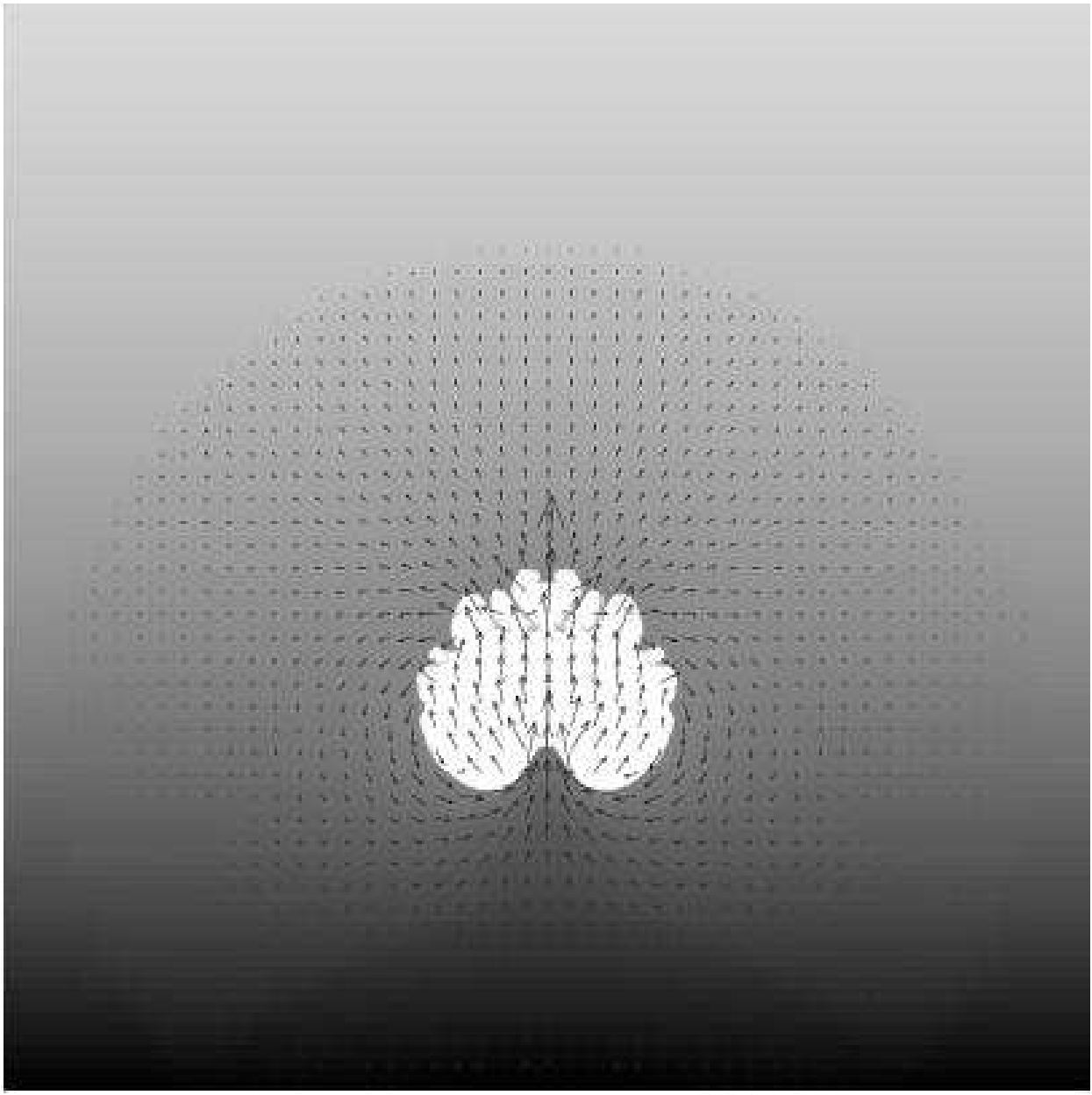}
\caption{A plot of the velocity field around the bubble in the fiducial case
(Fig.~\ref{fig:basecase}) at time $t \approx 72~\Myr$.
Of particular note is the two emerging large-scale rolls on
either side of the bubble, which will eventually form a stagnation line under the bubble,
generating the high-density `tail'.  }
\label{fig:basecase-externalflow}
\end{figure}

In the absence of magnetic effects, nothing prevents the complete
disruption of the bubble. The bubble's motion itself generates vortical
motion in the surrounding fluid of the same size as the bubble and with
rise velocity of the bubble, as shown in
Fig.~\ref{fig:basecase-externalflow}. These motions will then be able
to rip the bubble apart on order of a bubble rise time, $t_R = \sqrt{4
R / |g|}$, with $R$ being the bubble's radius, and $g$ being the local
gravitational acceleration.  These effects have been studied previously
(\emph{e.g.}, \citealt{bubblesreview} and \cite{layzer55}).

The bubble will also be susceptible to growth of the Rayleigh-Taylor
instability at the top.  Since we are dealing with large density
contrasts, the Rayleigh-Taylor growth time (from linear theory) will be
$t_{RT} = 1/\sqrt{kg} = \sqrt{ R / (2 \pi g n)}$ for the mode that is
$1/n$ the size of the bubble; however, modes of size on order the
bubble radius, or small modes near the sides of the bubble will be
suppressed because of the geometry of the bubble.  Comparing time
scales, the Rayleigh-Taylor instability will disrupt the bubble on
small scales near the center of the bubble quickly, but the disruption
of the entire bubble will be caused by induced vortical motions in the
fluid.  The shear along the side of the bubble will also be unstable to
the Kelvin-Helmholtz instability; this will generate secondary
instabilities as the evolution progresses.

The large scale motions of the fluid can be seen in
Fig.~\ref{fig:basecase-externalflow} at time $t \sim 70~\Myr$.
Two large rolls extend out, and their convergence
under the bubble results in an up-welling and compression of material,
forming a slight high-density tail in the wake of the bubble.

\begin{figure}
\plotone{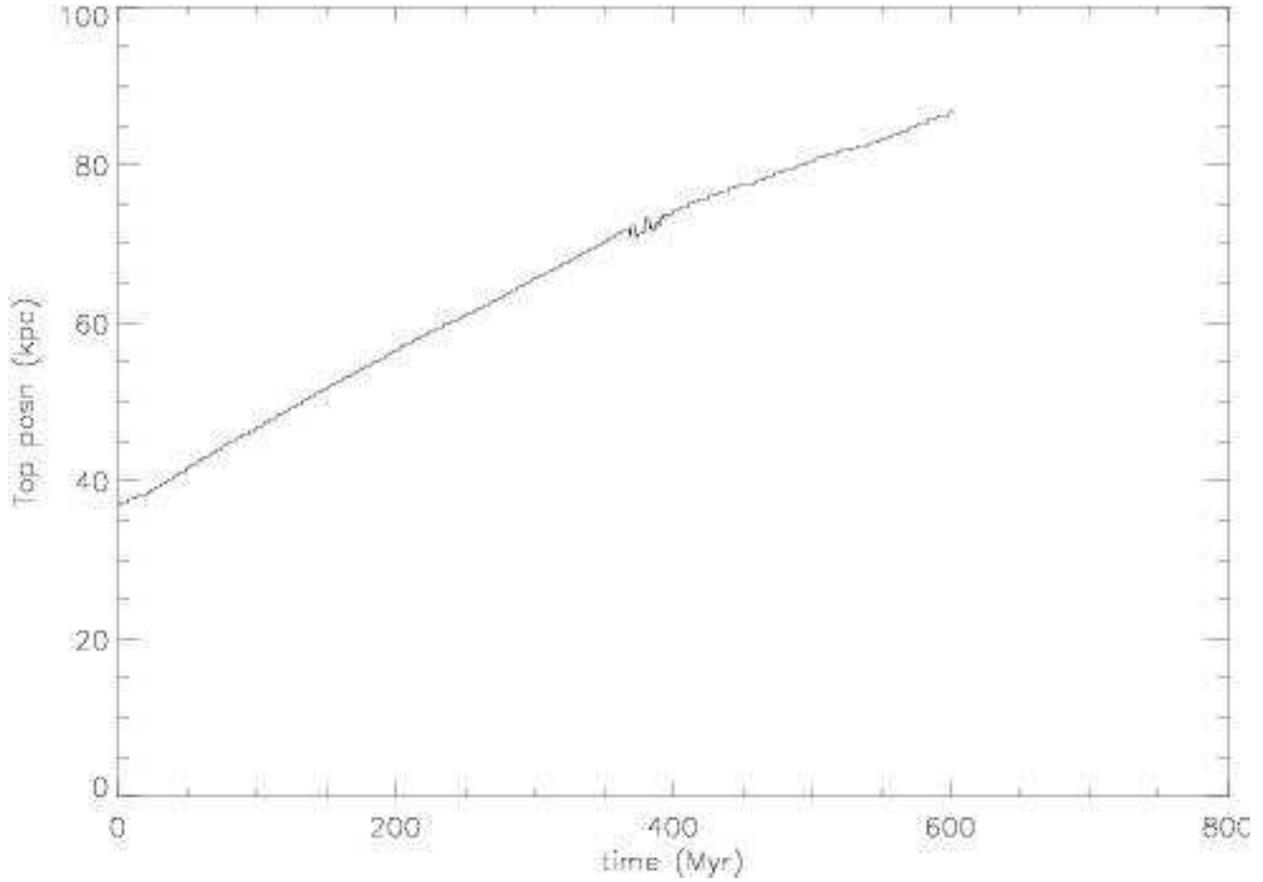}
\caption{The position of the top of the rising bubble in our reference
bubble case as shown in Fig.~\ref{fig:basecase}.  After an initial
acceleration, the bubble begins to slow down as it rises as both its
relative buoyancy changes and it entrains heavier material from the
environment.   The irregularity at $t \approx 375 \Myr$ is due to
vortical motions pulling down the clump that had been the head of the
bubble; some indication of this is shown in the middle frame of
Fig.~\ref{fig:basecase}.
}
\label{fig:basecasespeed}
\end{figure}

The position of the bubble top as a function of time is shown in
Fig.~\ref{fig:basecasespeed}.  This position was calculated by
horizontally averaging density throughout the domain and finding the
first position from the top of the domain of a density jump; this local
drop in density represents the presence of an under-dense fluid.
Because our underdensities were so large, the position measured this
way was insensitive to the precise value of the threshold we chose to
mark the density jump.  This measure allows us to track the top of a
`bubble' even when no coherent bubble exists.   The bubble slows down
over time primarily due to the entrainment of material from the
environment as the bubble is disrupted.

\begin{figure}
\plotone{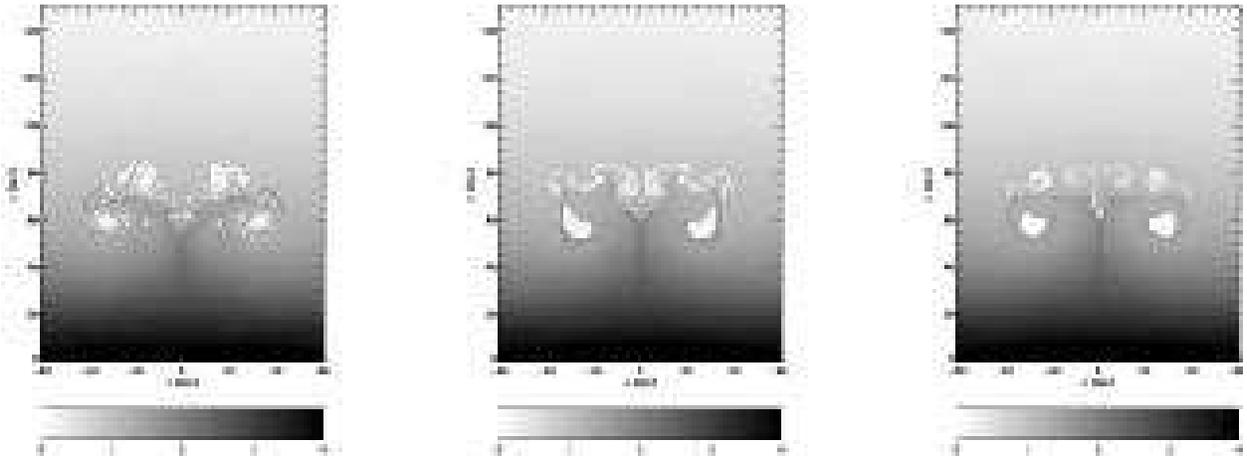}
\caption{The state of the bubble at $t \approx 580~\Myr$
         as resolution is decreased; at right, the reference case;
         middle, with resolution decreased to 3/8 of the full resolution ($192\times240$);
         right, with resolution decreased to 3/16 ($96\times120$).   Plotted is density.}
\label{fig:resolution-final}
\end{figure}

\begin{figure}
\plotone{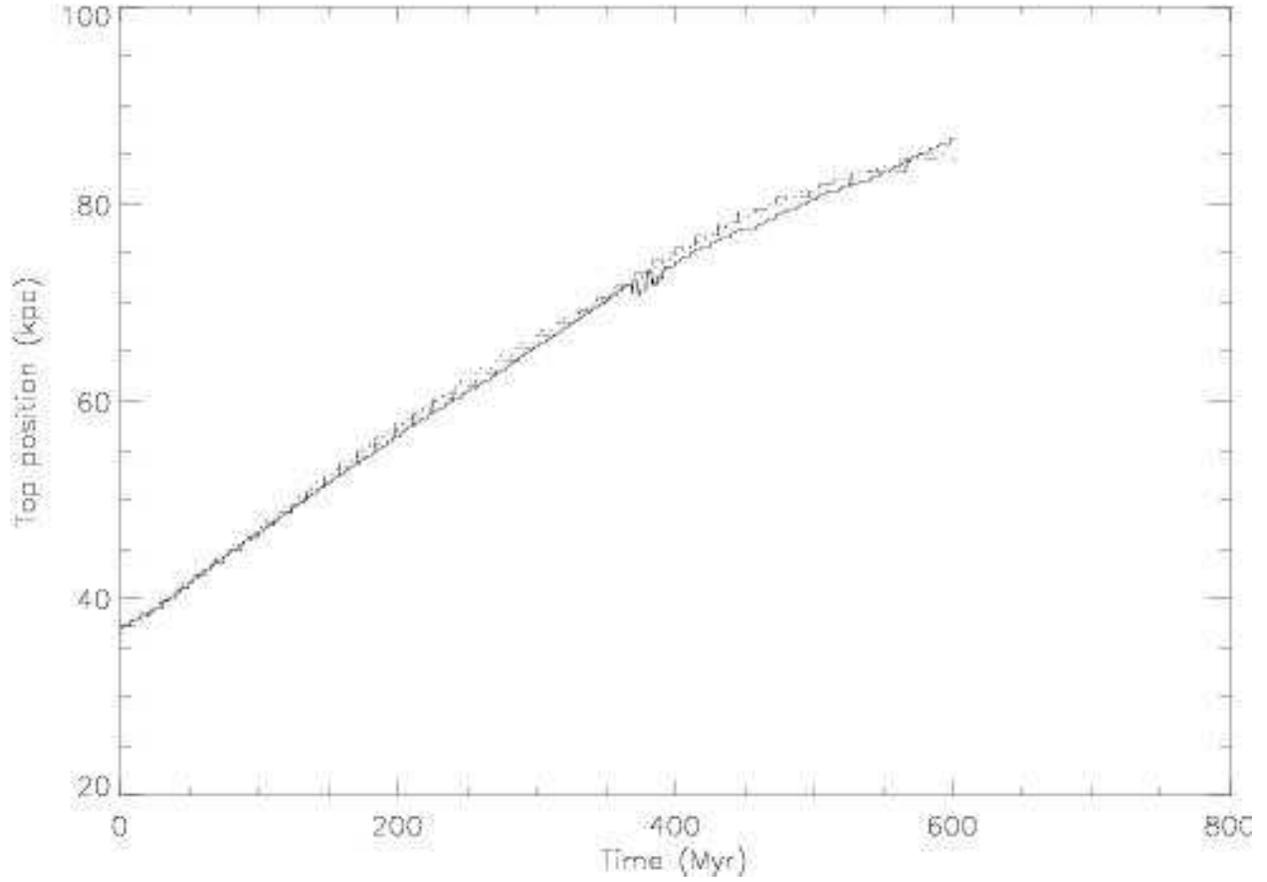}
\caption{The position of the top of a $r = 9 \kpc$ bubble vs time for
         three resolutions.  Solid line the resolution used in this
         study; dotted line decreased to a factor of 3/8; and 
         dashed line decreased to a factor of 3/16.}
\label{fig:resolution-vstime}
\end{figure}

As resolution varies, the small-scale structure varies, but the overall
behavior remains the same, as is shown in
Fig.~\ref{fig:resolution-final} and Fig.~\ref{fig:resolution-vstime}.
In these simulations, there is no modeled dissipation mechanism such
as viscosity or thermal or material diffusion.    In this case, only
the numerical dissipation on small scales limits the amount of small
scale structure.  The small scale structure affects entrainment of
ambient (denser) material, so that as resolution increases and
numerical dissipation decreases, late-time heights change, but the
early behavior is nearly identical.

A typical rise velocity for the still largely intact bubble can be seen
from looking at the early-time slope in Fig.~\ref{fig:basecasespeed};
it is approximately $0.1~\kpc~\Myr^{-1}$.   That this is near the sonic
velocity  ($0.41~\kpc~\Myr^{-1}$) is not a coincidence.  The density
contrasts we consider here are large enough that the buoyant
acceleration felt by the bubble is approximately $g$; since in an
isothermal atmosphere the scale height is $H = c_s^2/(\gamma g)$, where
$c_s$ is the sound speed, the velocity that the bubble picks up as it
rises a significant fraction of a scale height is on the order $c_s$.
The disruption of the bubble and entrainment of denser material before
a scale height is reached prevents the bubble from accelerating past the 
sound speed.


\subsection{Effects of Density Contrast}
\begin{figure}
\plotone{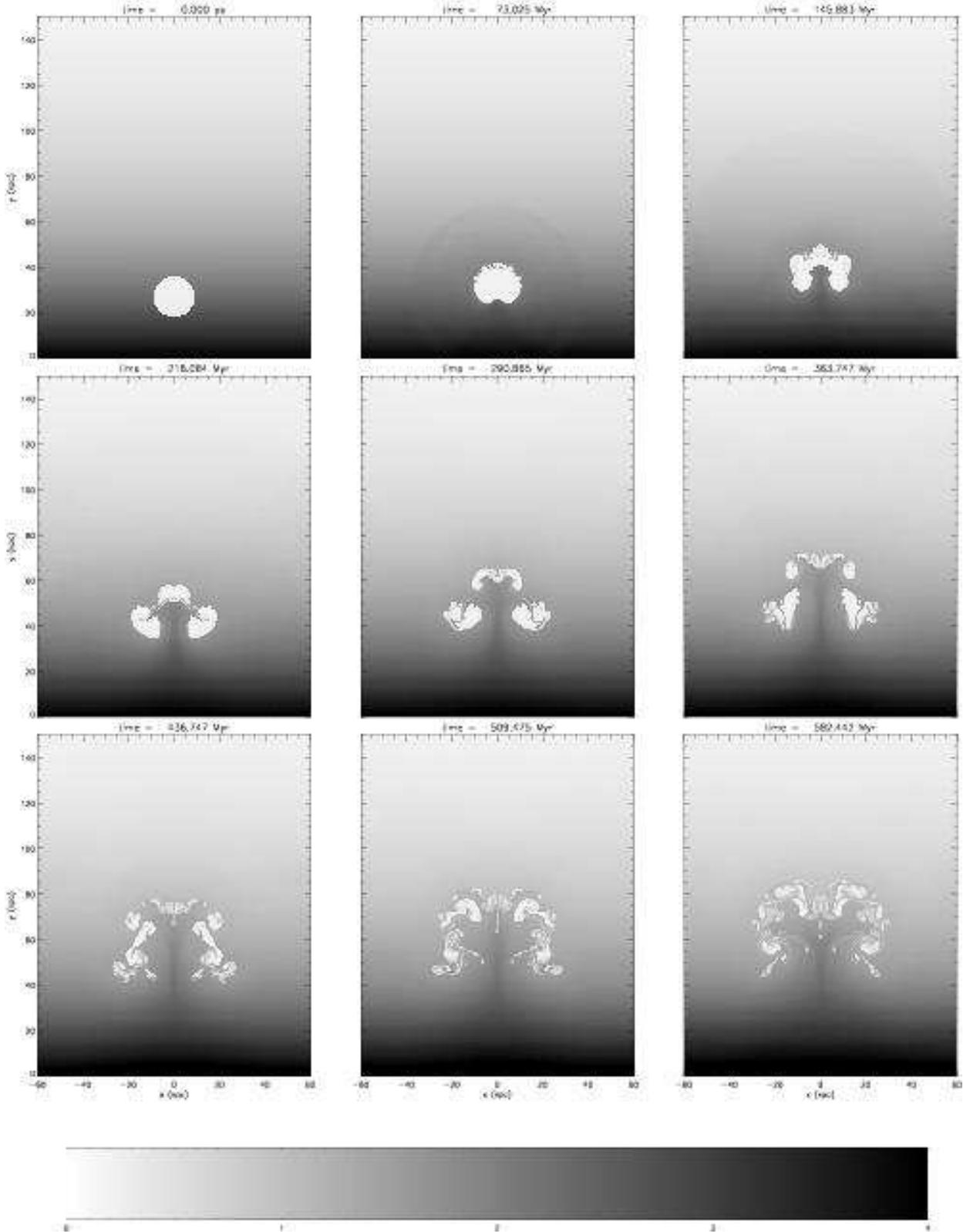}
\caption{The time evolution of the 10:1 density contrast case.   Shown in the
panels is density.}
\label{fig:lodenscase}
\end{figure}

We ran simulations with density contrasts between the bubble and the
maximum density in the box of 100:1 and 10:1.  While there are
some differences between the resulting structures (compare
Figs.\ref{fig:basecase} and \ref{fig:lodenscase}), caused partly by
the different growth of the Rayleigh-Taylor instability at the top of
the bubble, the underlying dynamics are broadly similar.   One way of
understanding this is that in both cases, the bubble densities are very
much less than the surrounding environment, so that it is the momentum
of the fluid in the environment, not the bubble, which dominates the
dynamics.  Quantitatively, the Atwood number, $A =
(\rho_1-\rho_2)/(\rho_1+\rho_2)$ is $0.82$ in the 10:1 case, and $0.98$
in the 100:1 case -- so while the bubble density changes by a factor of
10, the Atwood number, which controls both the buoyant force and the
evolution of the Rayleigh-Taylor instability, changes only by 20\%.

\begin{figure}
\plotone{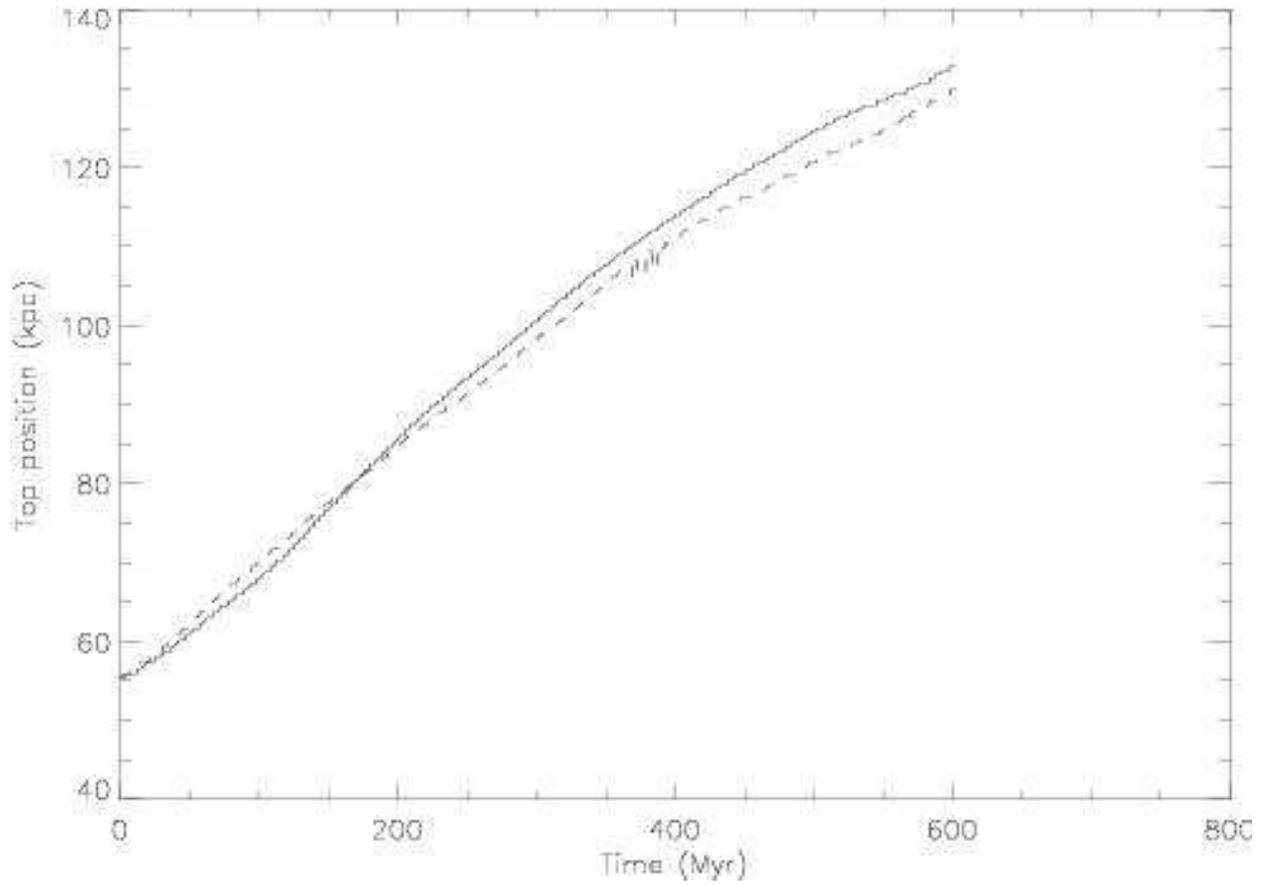}
\caption{The position of the top of a rising bubble as shown in
Fig.~\ref{fig:basecasespeed} (dashed line) and with the same parameters
except a lower, 10:1, density contrast (solid line).}
\label{fig:lodensspeed}
\end{figure}

In particular, looking at the position of the top of the bubble versus
time (in Figs.~\ref{fig:basecasespeed} and
\ref{fig:lodensspeed}) we see a very similar evolution after the
initial acceleration phase.  Perhaps surprisingly, the bubble with the
larger density (the 10:1 case) actually rose slightly further during
the course of the simulation.  In the late-time evolution, it is the
entrainment of higher-density fluid which determines the buoyancy, and
thus the acceleration, of the bubble.   The bubble in the 10:1 case
entrained less ambient fluid, thus its buoyancy ended up being greater
than in the nominally more buoyant 100:1 case.


\subsection{Effects of Geometry}
\begin{figure}
\plotone{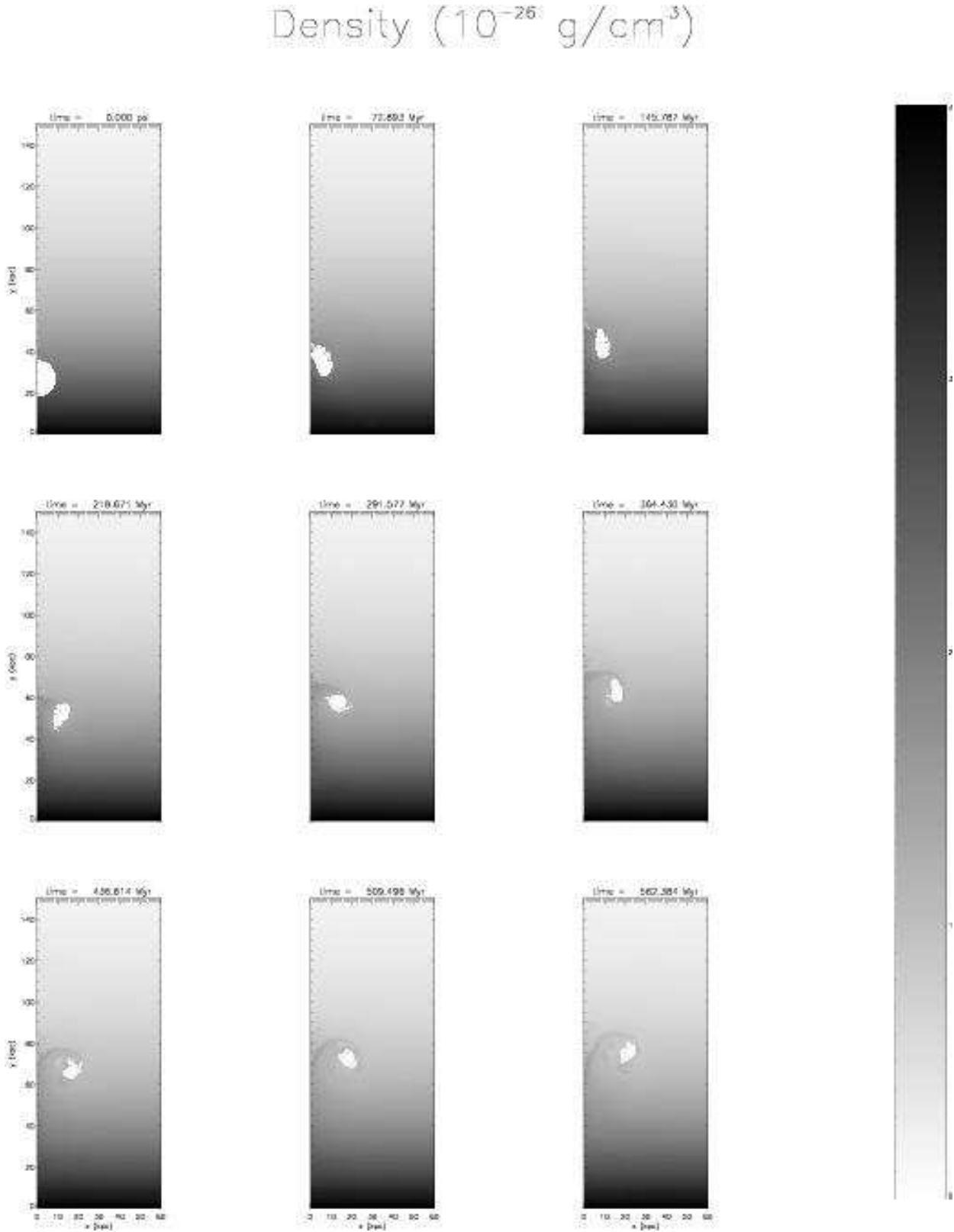}
\caption{The evolution of a 9 \kpc~radius bubble in a 100:1 density contrast in cylindrical coordinates.  The bubble is torn into two tori with a train of higher density material.   Plotted is density.}
\label{fig:cylind}
\end{figure}

In the purely hydrodynamic case it is possible to perform the same
simulations described above in 2-D axisymmetric cylindrical
coordinates, making the bubble a sphere instead of a cylinder.  In this
more physically relevant case, the bubble deforms into
tori (Fig.~\ref{fig:cylind}), rather than the wing-like shape found in
planar coordinates, as the shedded vorticity from the bubble rising
becomes a vortex ring rather than two parallel lines of vorticity.

Because in this geometry there is less volume near the center of the
bubble than there is towards the edges, the vortices that form can
advect essentially all of the material from within the bubble into the
vortex structure, so that the bubble becomes a rising vortex ring which
then undergoes secondary instabilities.  Also because of the geometry,
the effect of the initial Rayleigh-Taylor instability in disrupting the
bubble is diminished, although it does not vanish; the material that
is disrupted by the Rayleigh-Taylor instability forms a smaller,
secondary vortex ring that leads the larger ring.

\begin{figure}
\plottwo{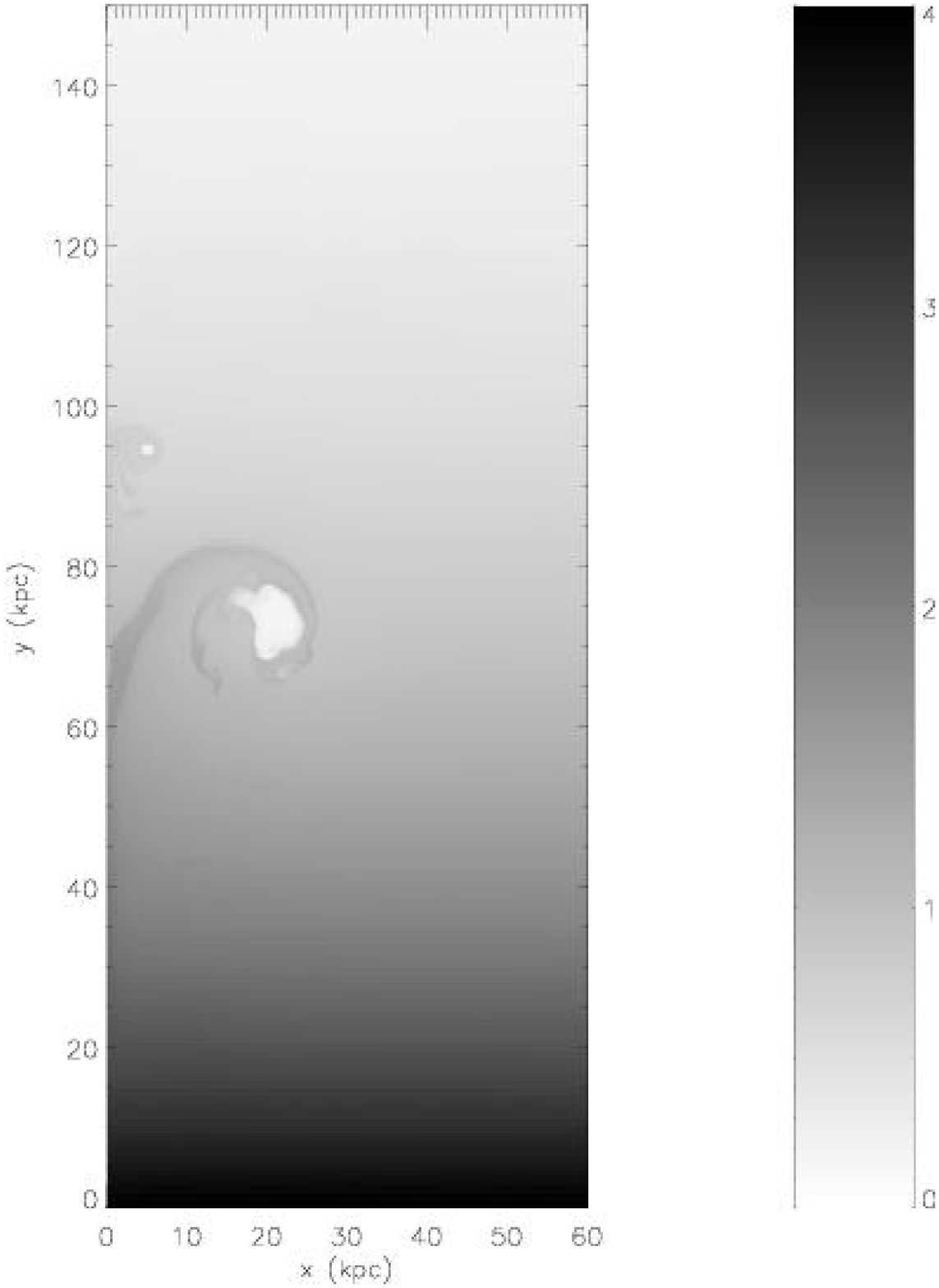}{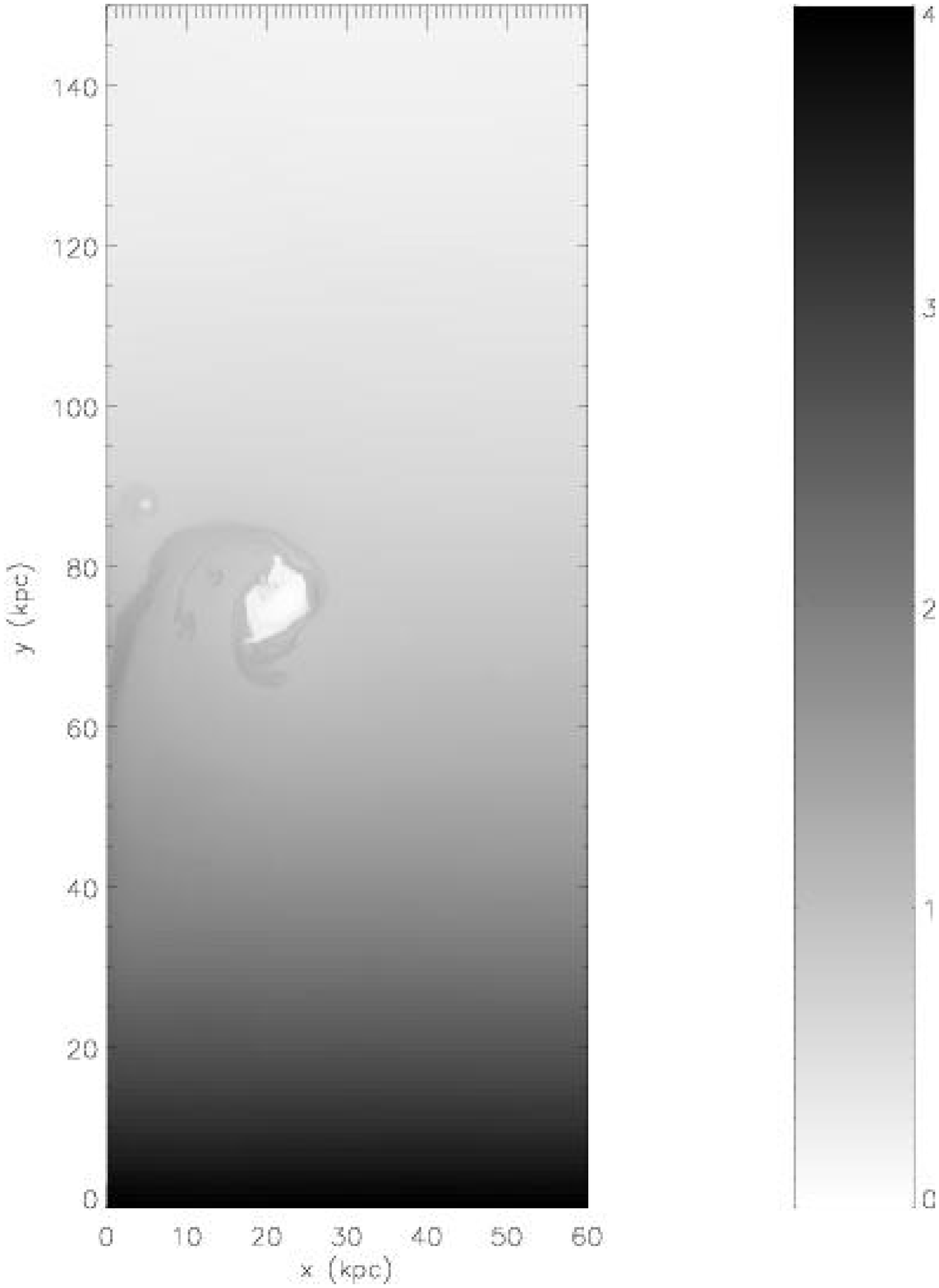}
\caption{The effect of differing density contrasts for the simulations in cylindrical
symmetry. Left, 10:1; right, 100:1.  Frames are taken at time $t \approx 580~\Myr$.
Density is plotted.}
\label{fig:cylind-contrast}
\end{figure}

This simulation was performed with several density differences which
all exhibited similar behavior.  Fig.~\ref{fig:cylind} shows the
evolution of one of these cylindrical cases with a density contrast of
100:1 and a 9 \kpc \ radius bubble.  Fig.~\ref{fig:cylind-contrast}
shows the final frame of this simulation along with the
corresponding frame of a simulation performed with a 10:1 density contrast.
As with the planar case, although there are differences in the
small-scale structure, the medium- and large-scale behavior is
essentially unchanged.


\subsection{Effects of Bubble Size}
\begin{figure}
\plotone{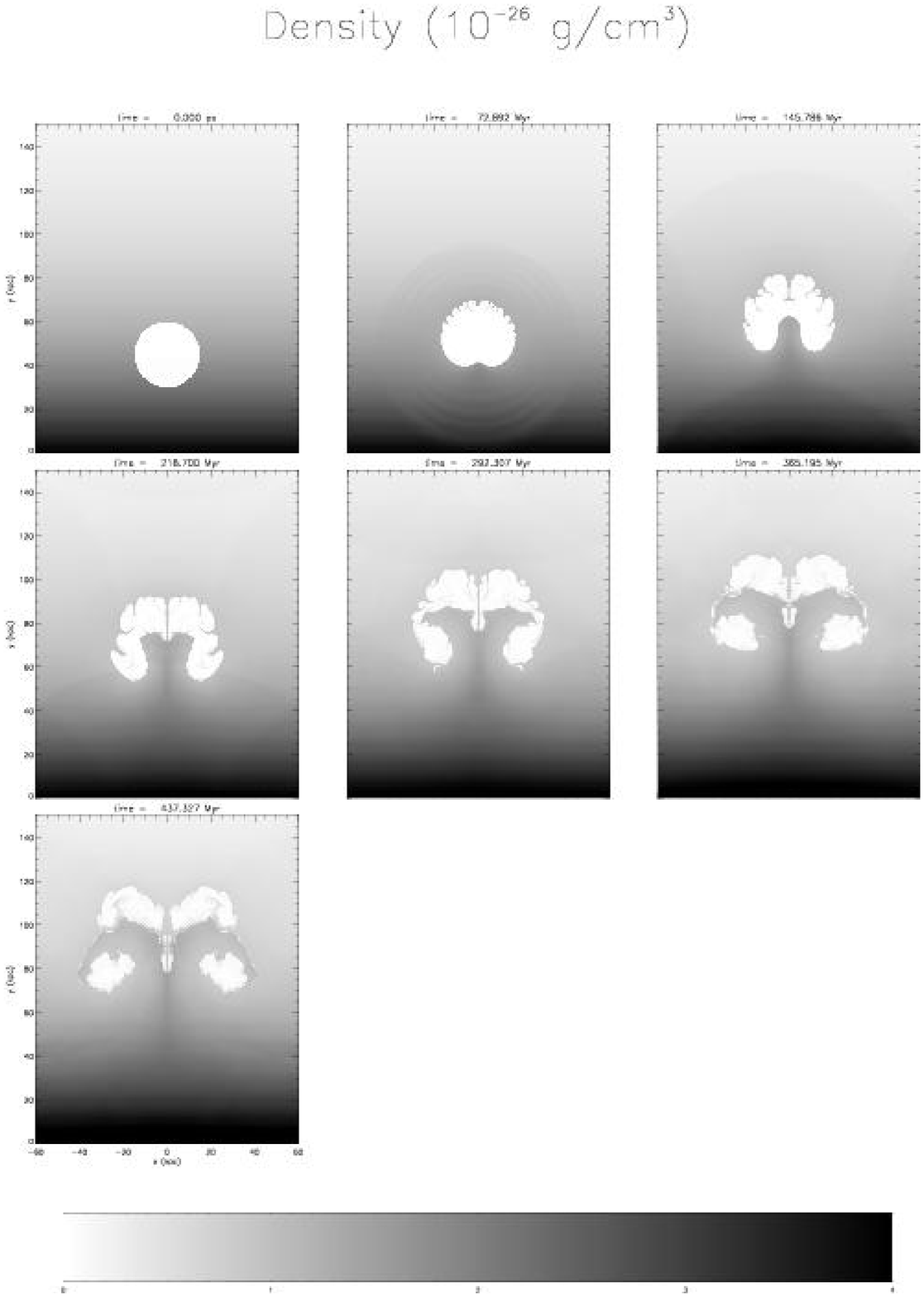}
\caption{The evolution of a $r = 15~\kpc$ bubble in a 100:1 density contrast, $g = -7 \times 10^{-9} \cmss$ atmosphere.  The Raleigh-Taylor instability grows much further
in this case, as there are more modes to grow on the larger bubble.  
The larger growth rate means that significant evolution has occurred
before the induced vortical motions begin to rip the bubble apart in
earnest; thus there is a much larger `head' to the structure than in
the base case in Fig.~\ref{fig:basecase}.   The simulation was ended
early as the bubble approached the boundaries of the domain.  Density is plotted.}
\label{fig:bigbubble}
\end{figure}

\begin{figure}
\plotone{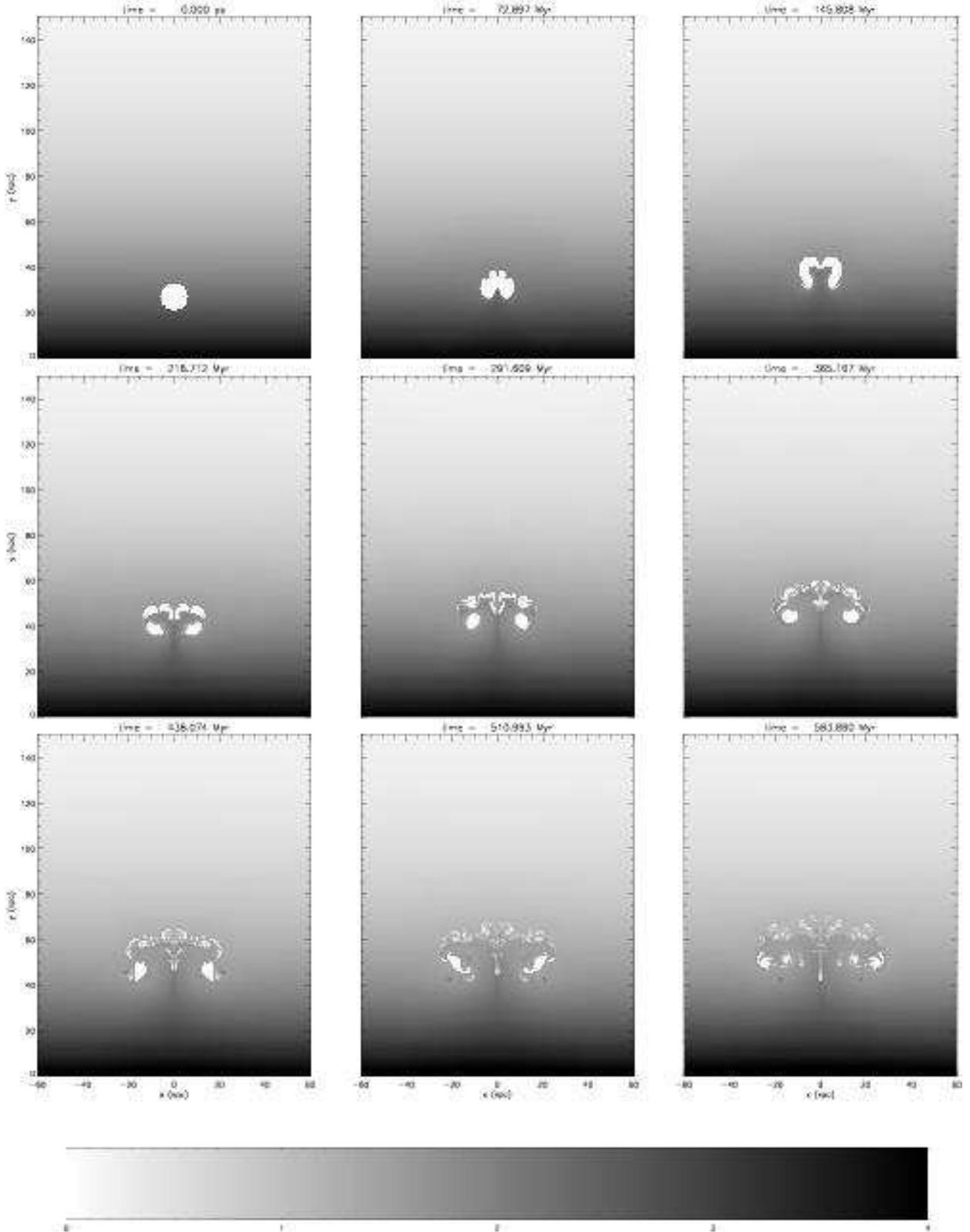}
\caption{The evolution of a $r = 6~\kpc$ bubble in the 100:1 density contrast, $g = -7 \times 10^{-9} \cmss$ case.
Plotted is density.}
\label{fig:smallbubble}
\end{figure}

Simulations were also performed with a larger and a smaller bubble in the
planar case.  In the case of the larger $r = 15\ \kpc$ bubble of
Fig.~\ref{fig:bigbubble}, the Raleigh-Taylor instability plays a much
larger role in the evolution of the bubble than in the previous case,
as there are more modes that can grow on the larger top of the bubble.
Conversely, the $r = 10\ \kpc$ bubble (Fig.~\ref{fig:smallbubble})
tends to remain flat across the top, and is primarily disrupted by the
induced vortical motions.


\subsection{Effects of Stratification}
The isothermal atmosphere surrounding the bubble has an exponential
density profile provided by a constant gravitational field in the
negative $y$-direction.  Changing the stratification, through the
gravitational acceleration, has two major effects.  Immediately, it
changes the time scale on which the dynamics occurs, as shown in the
scaling relations in \S\ref{sec:initialmodels}.  The stratification
also tends to elongate vortical motions in the atmosphere, due to lower
densities at higher positions.

\begin{figure}
\plotone{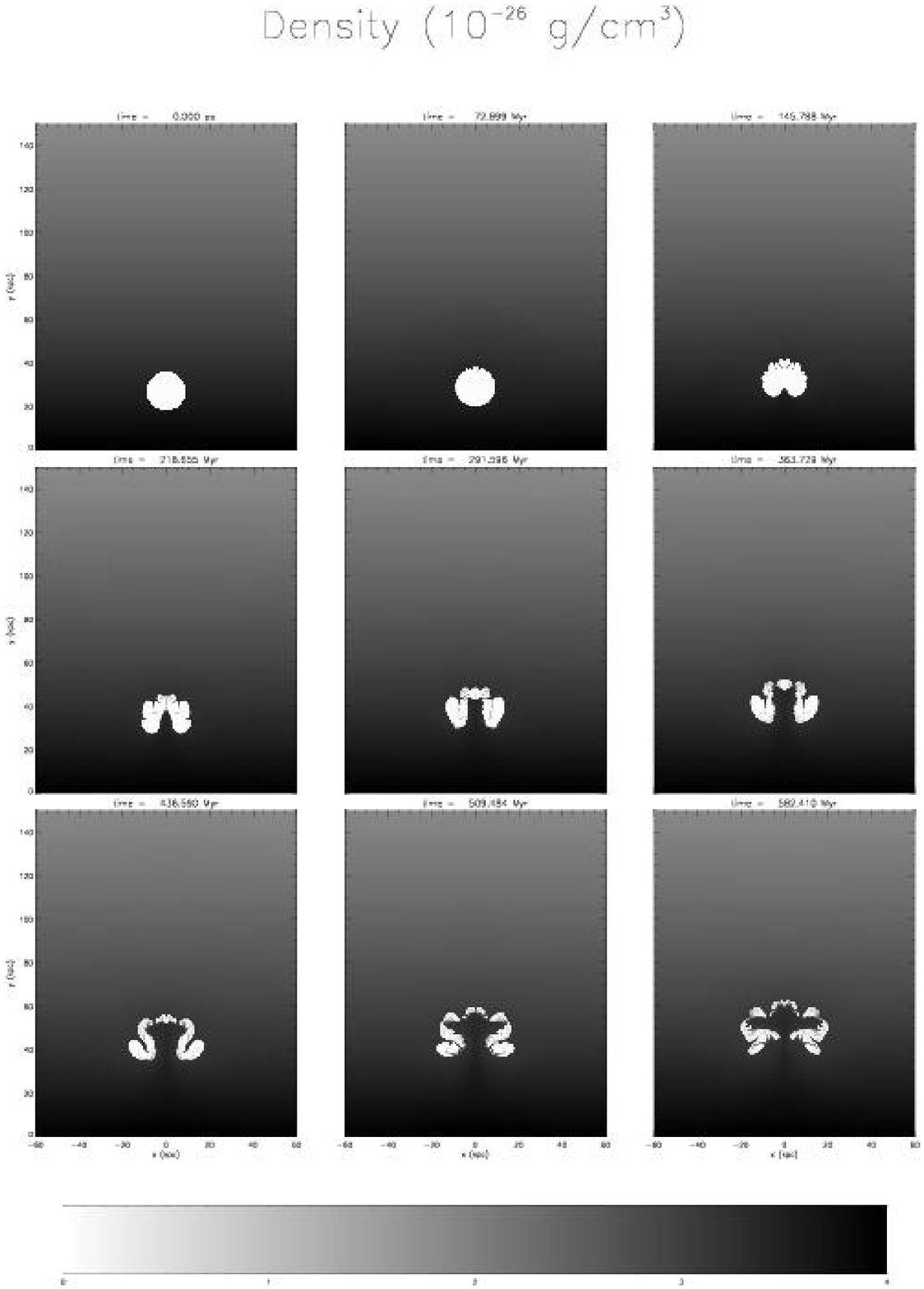}
\caption{The evolution of a $r = 9~\kpc$ bubble in a less-stratified
         atmosphere, with $g = -1.75\times 10^{-9}~\cmss$.  Plotted is density.}
\label{fig:lowgrav}
\end{figure}

\begin{figure}
\plottwo{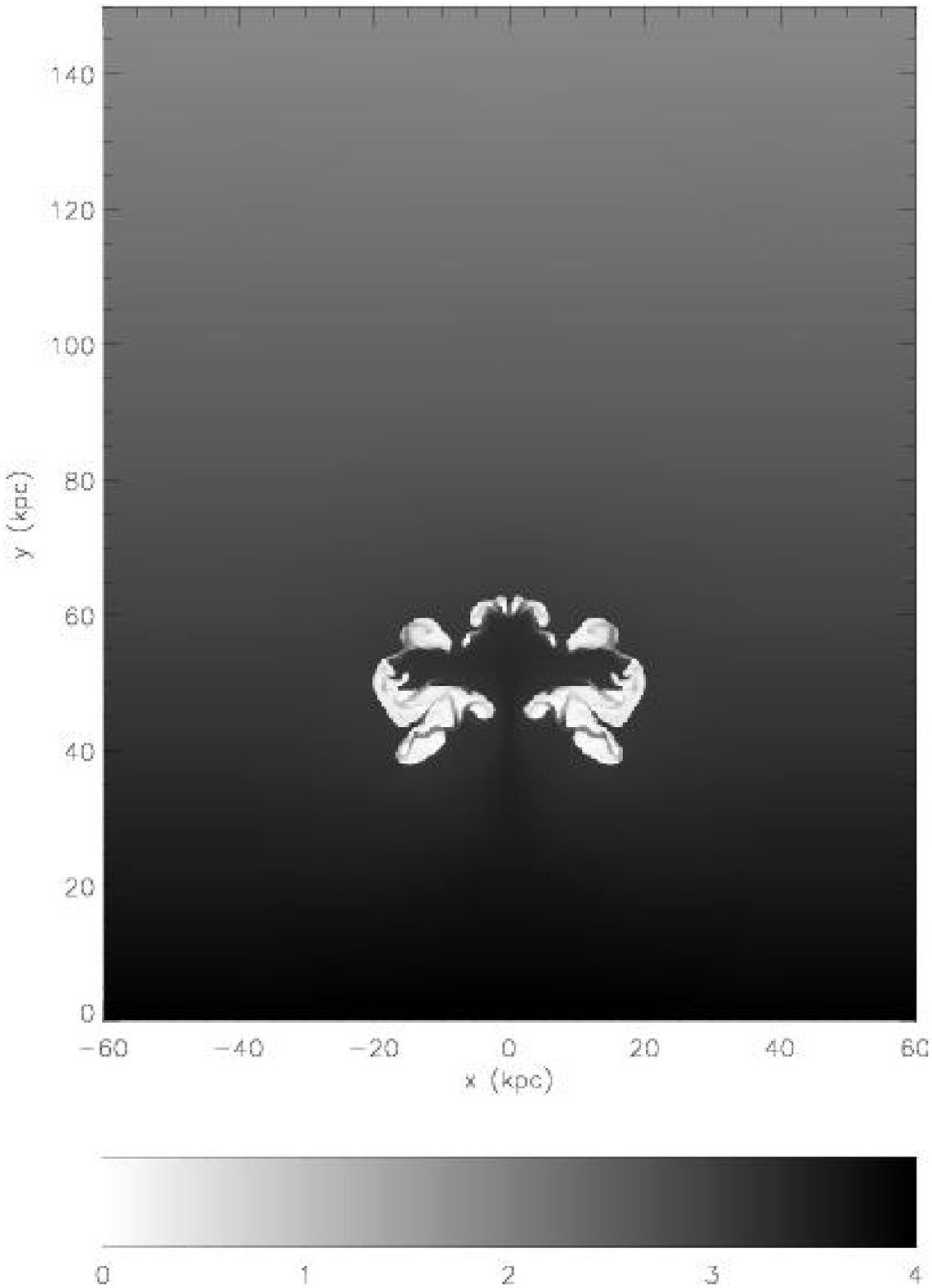}{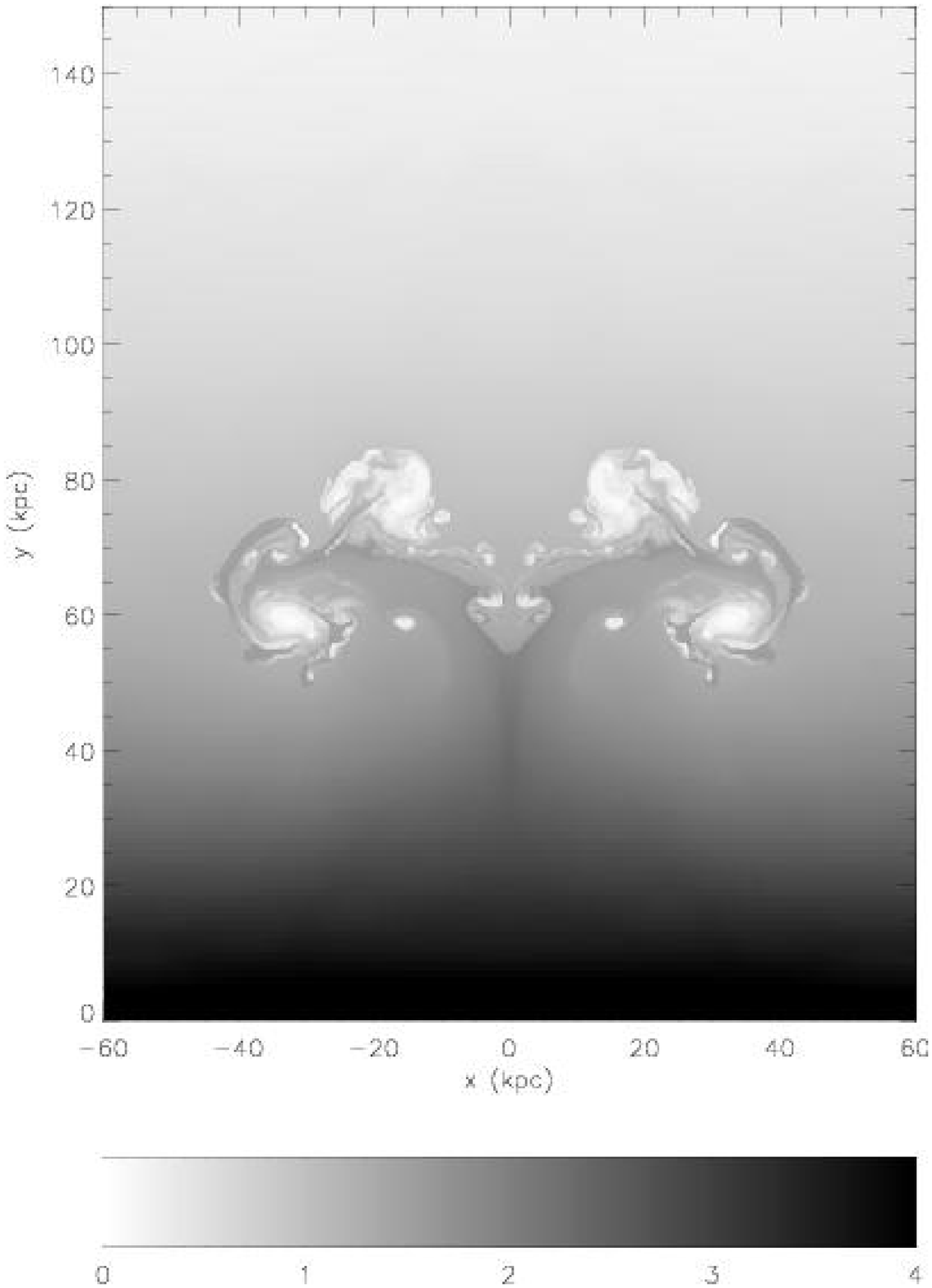}
\caption{The effect of stratification on the evolution of the
bubble; shown at left is the density field for the simulation with
$g=-1.75\times 10^{-9} \cmss$ at $t \approx 580\ {\mathrm{Myr}}$, and on the right,
that for $g = -7\times 10^{-9} \cmss$ at $t \approx 290\ {\mathrm{Myr}}$.  Plotted
is density.}
\label{fig:strat}
\end{figure}

In Fig.~\ref{fig:lowgrav} we show the evolution of a bubble moving
through a less stratified domain, with the gravity set to be one
quarter of its value in our reference case.   In Fig.~\ref{fig:strat},
we compare the final frame of the evolution with the frame from the
corresponding time (taking into account the change of the buoyant
time scale) of our reference case.

\begin{figure}
\plottwo{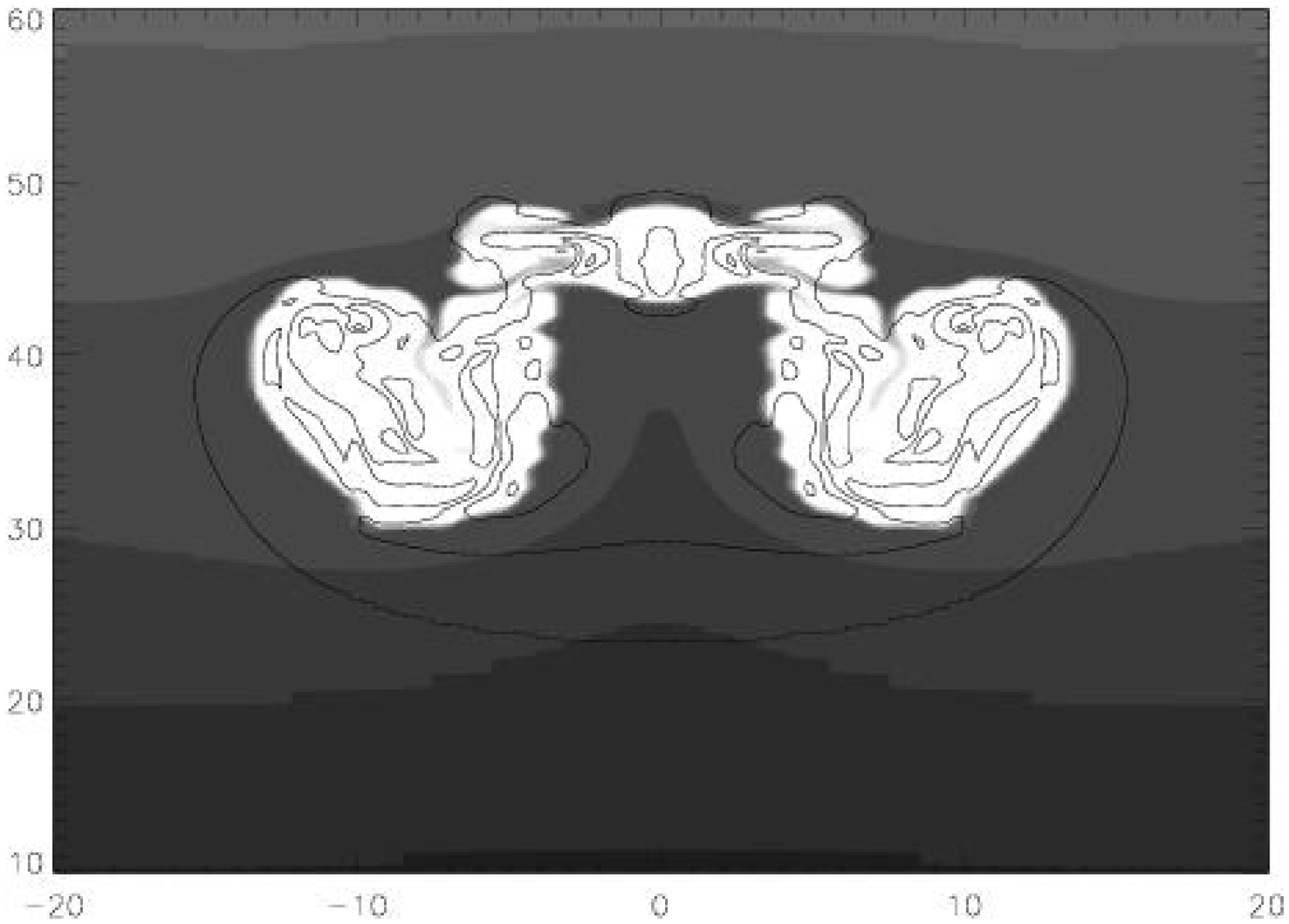}{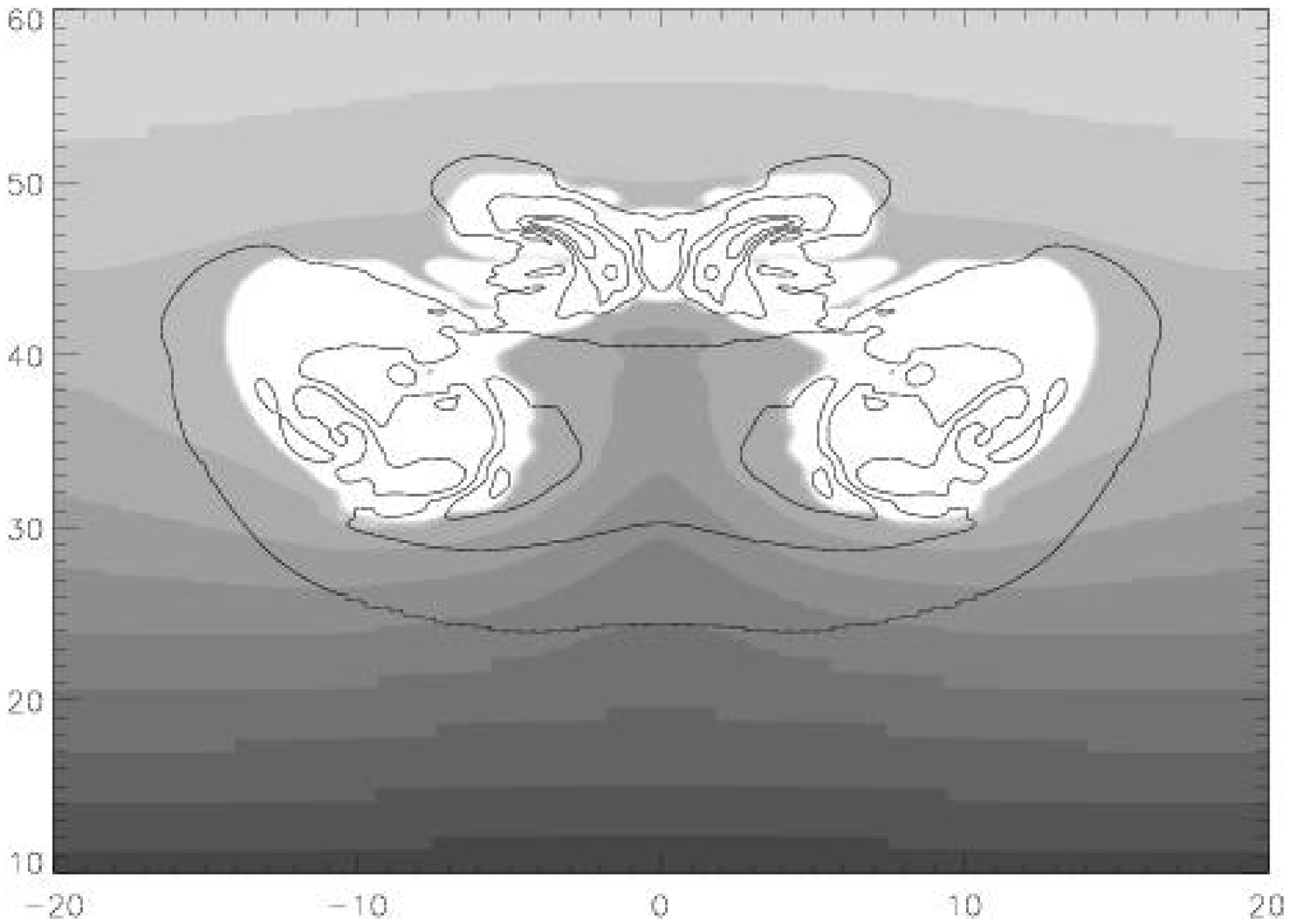}
\caption{Scaled total velocity contours superimposed on the density
field; shown at left are results for the simulation with
$g=-1.75\times 10^{-9} \cmss$ at $t \approx 290\ {\mathrm{Myr}}$, and on the right,
that for $g = -7\times 10^{-9} \cmss$ at $t \approx 145\ {\mathrm{Myr}}$.
In the more strongly stratified case (on the right), the velocity
contours extend further up than in the less stratified case.
}
\label{fig:stratvel}
\end{figure}

Some of the differences between the two simulations can be understood
in terms of the height of the vortical motions induced by the rising
bubble.  The large-scale motions, if they are to conserve momentum,
must be elongated in atmospheres where the density is stratified, as
there is less mass at higher positions in the atmosphere.
Fig.~\ref{fig:stratvel} shows total velocity contours on top of a
zoom-in of density for the same simulations as in Fig.~\ref{fig:strat},
but at earlier times.    The vortical motions extend noticeably higher
in the more stratified case, meaning that in this case the bubble will
be disrupted over longer length scales.

\section{MAGNETIC FIELDS: HOT BUBBLE IN A BACKGROUND FIELD}
\label{sec:magneticfields}
From purely hydrodynamic bubbles, we move on to examining hot bubbles
rising in a fluid with an initially constant background field running
through the entire domain.   In the rest of this paper, we describe
solutions performed with the MHD solver in the \FLASH{} code.   The rest
of the code, including the boundary conditions and initialization,
remained unchanged; the MHD solver was simply compiled into the
simulation code in place of the PPM solver.  This MHD solver is based
on the code described in \cite{mhdcodepaper}, with two methods of
`$\nabla\cdot {\bf{B}}$-cleaning' -- a diffusive method due to
\cite{diffusecleaning}, and a projection method due to
\cite{projectcleaning}.  We used the diffusive mechanism for the
results described here due to the more modest computational cost of
this method and the simple magnetic dynamics in these simulations.
Our implementation of this MHD solver works only in Cartesian
coordinates, so we consider only the 2-D planar cases.

\begin{figure}
\plotone{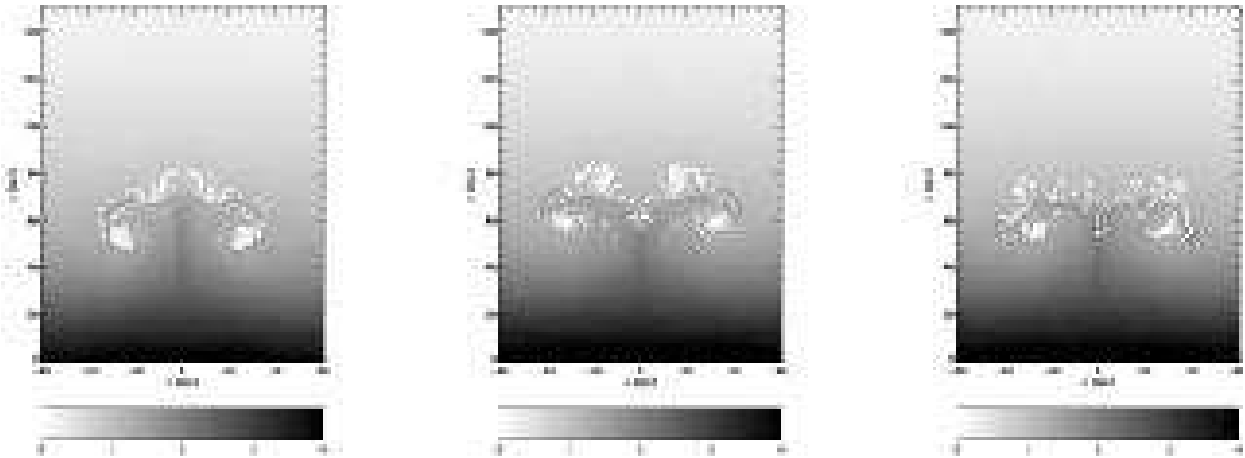}
\caption{
Left, the small scale structure at 580 \Myr{}  using the MHD solver
with no magnetic field.  Center, the small scale structure at the same
time calculated with the PPM solver.   Right, the small scale structure
at the same time using the PPM solver with linear, rather than
parabolic, reconstruction functions more similar to that used by the MHD solver.
Plotted is density.
}
\label{fig:basemhd}
\end{figure}

Since we are changing the numerical solver as well as adding magnetic
fields to the initial conditions, it is important to ensure that
we get the same results in the purely hydrodynamic case with both
solvers.  Calculating our reference case without magnetic
fields with both solvers, we find noticeable, but understandable,
differences in the final results.   Shown in Fig.~\ref{fig:basemhd}
is our reference calculation at the final time calculated with our
MHD solver and the same result with the PPM solver.  The MHD result
looks something like the lower-resolution versions of the PPM solution
(Fig.~\ref{fig:resolution-final}), which can be understood to be
a consequence of the the lower spatial order of accuracy of the MHD solver.
Fig.~\ref{fig:basemhd} also shows the results of using the purely
hydrodynamic solver with reconstruction functions more similar to those
in the MHD solver, which improves the agreement between the
simulations.  Understanding the differences between the results of the
different solvers gives us some confidence that the results obtained
with the magnetic field solver can be meaningfully compared with those
calculated using the purely hydrodynamic solver.

In the results that follow, a density ratio of 10:1 was used.  This
avoids numerical issues which can arise if there is too large a jump in
Alfv\'en velocity across a small region (eg., between the bubble and
the ambient material.)

\subsection{Horizontal Field}

\subsubsection{z-field}

The simplest case to consider in 2-D is an initial magnetic field in
the $\bf{\hat{z}}$-direction, out of the plane of the simulation domain.
In this case, the magnetic field lines can slip around the cylinder as it
rises, enhancing the initial production of large-scale vorticity around
the bubble.   The magnetic field's contribution to the dynamics in this
case is simply as an additional, constant, pressure term $B^2/(8 \pi)$,
reducing the effective relative pressure stratification $\nabla p / p$
by a factor of $\beta_0/(\beta_0 + 1)$ while leaving the density
structure unchanged.  Here $\beta_0$ is the plasma $\beta$ parameter
(gas pressure divided by magnetic pressure) at the base of the
atmosphere.  This effective pressure scale height will be one
characteristic length scale for eddies, so that decreasing $\beta$
will, up to a limit, decrease the characteristic vortical motion size
without the elongation that increasing the density stratification would
produce.

\begin{figure}
\plotone{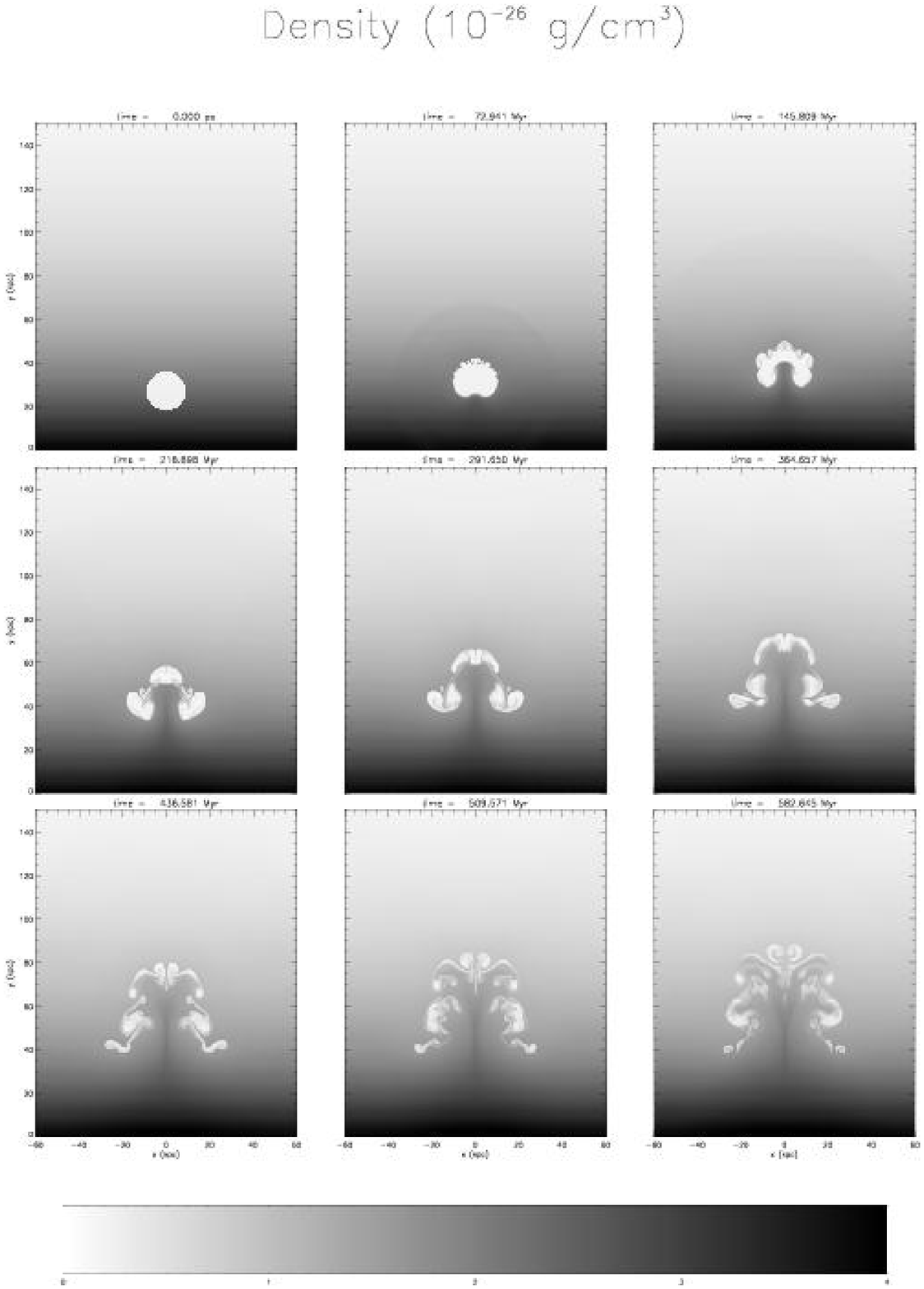}
\caption{The time evolution of a 10:1 bubble rising in a constant
initial background magnetic field in the $\bf{\hat{z}}$-direction (out of the
plane of the paper) with $\beta_0 = 462$.   Plotted is density.}
\label{fig:zmagevolve004}
\end{figure}

\begin{figure}
\plotone{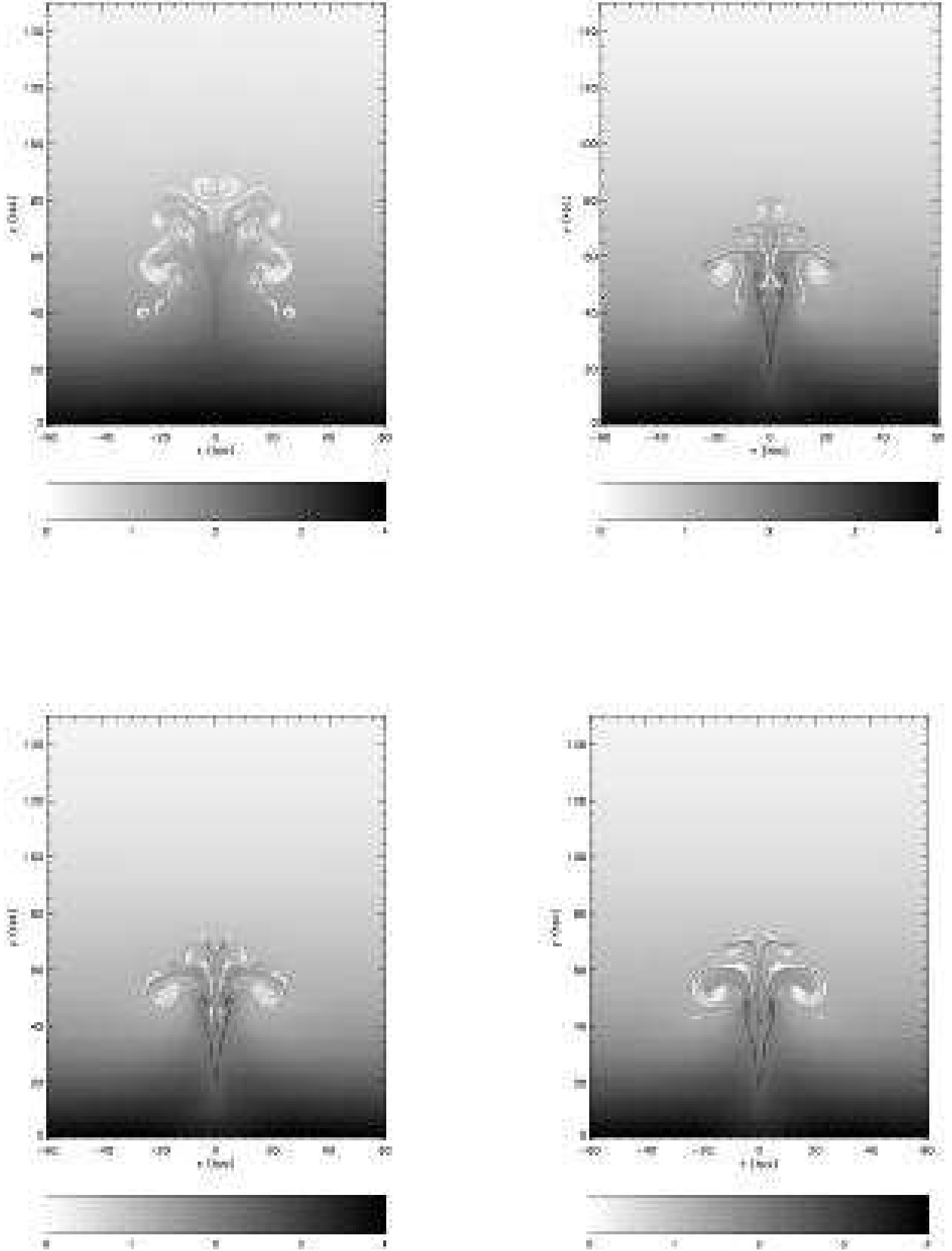}
\caption{The state at time $t \approx 580~\Myr$ of a bubble rising in 
a medium with a constant magnetic field in the $\bf{\hat{z}}$ direction
with $\beta_0= 462$ (top left), $0.185$ (top right), $0.046$ (bottom left),
and $0.012$ (bottom right).  The panels are plots of density.}
\label{fig:zhatall}
\end{figure}

The evolution of a bubble in a very weak $\bf{\hat{z}}$-field is shown
in Fig.~\ref{fig:zmagevolve004}.  In Fig.~\ref{fig:zhatall} is shown
the final frame for simulations with $\beta_0$ of 462, 0.19, 0.046, and
0.012.     As $\beta_0$ decreases, the vortical structures are seen to
`tighten'.

At late times, the high-density wake that the bubble leaves begins
to contain low-density structures; one explanation for this is
that the large-scale motions induced by the flow sweep up magnetic
field lines, resulting in a higher magnetic field density in the wake
of the bubble; the resulting increase in magnetic pressure means a
lower mass density under these constant-pressure conditions, leaving a
different tail in this case than in the purely hydrodynamic case.

\subsubsection{x-field}

As opposed to the effects of a field in the $\bf{\hat{z}}$-direction,
even an extremely weak field in the $\bf{\hat{x}}$ direction in this
geometry can strongly contain the rising bubble in our planar
coordinates.   We see from \cite{chandra} that the magnetic
field suppresses the Kelvin-Helmholtz instability completely
unless the relative velocity between the bubble and the ambient
medium exceeds the root-mean square Alfv\'en speed in the two media;
or, as written in \cite{magfield},
\begin{equation}
4\pi\frac{\rho_0\rho_b}{\rho_0 + \rho_b}v^2 > B^2_0 + B^2_b.
\label{eq:shearinstability}
\end{equation}
At the top of the bubble, near the stagnation point, the relative
velocity is nearly zero and of course the instability is suppressed.
Even near the sides, however, because the Alfv\'en speed in the bubble
is so high (due to the low density), the shear velocity would have to
be many times the ambient Alfv\'en speed to for Kelvin-Helmholtz
instabilities to occur.

\begin{figure}
\plotone{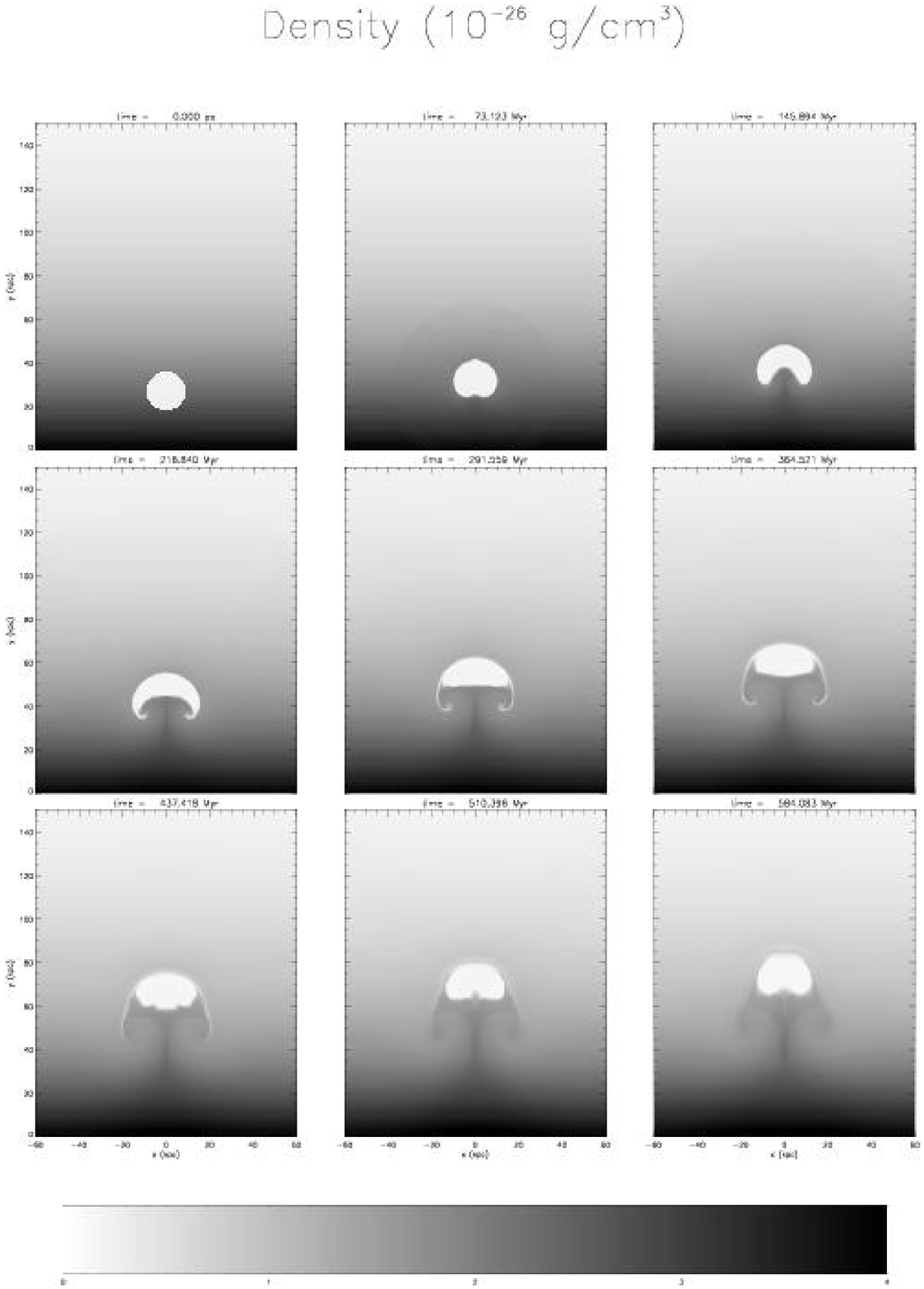}
\caption{The evolution of a bubble in a horizontally magnetized background plasma with magnetic field in the ${\bf{\hat{x}}}$-direction, $\beta_0 = 462$.   Plotted is density.}
\label{fig:magx}
\end{figure}

Initially, since $B_0 = B_b$, this reduces to $\beta_0 > (4/\gamma)
(\rho_0/\rho_b + 1) {\cal M}^{-2}$ where $\gamma$ is the ratio of
specific heats for the gas, and ${\cal M}$ is the flow Mach number.
Since we have a density ratio of 10:1, a Mach number of 0.25, and a
5/3-law ideal gas, this suggests that for $\beta_0 > 420$, the bubble
should be unstable to shear.   However, we show the evolution of a
bubble with $\beta_0 = 462$ in Fig.~\ref{fig:magx}, and there is
clearly no shear instability present.   In fact, as the bubble rises,
not only does shear occur but the magnetic field lines are wrapped
around the rising bubble, increasing the field density.  Further, the
shear at the bubble interface is overestimated by the net bubble
velocity upwards.  Applying Eq.~\ref{eq:shearinstability} to the data
of Fig.~\ref{fig:magx}, we find that the shear everywhere is several
orders of magnitude below the threshold for instability.  This is
tabulated in Table~\ref{tab:stabilize}.

In this case of a horizontal magnetic field, the wake left behind by
the bubble is an unstructured high-density entrained tail; this will be
true in the spherical bubble case, as it is caused by the stagnation
line caused by the induced vortical motions of the rising bubble.

The drastically different effects of the two horizontal orientations of
the uniform magnetic field is partly an artifact of the geometry of the
simulation.  For a spherical bubble, rather than a cylindrical bubble,
a uniform horizontal field would act on the bubble asymmetrically -- as
a tension in one plane, and as a pressure component in the other.
In particular, the strong containing effect of even a weak
$\bf{\hat{x}}$-field is a geometrical artifact of these simulations.

\subsection{Vertical Field}

\begin{figure}
\plotone{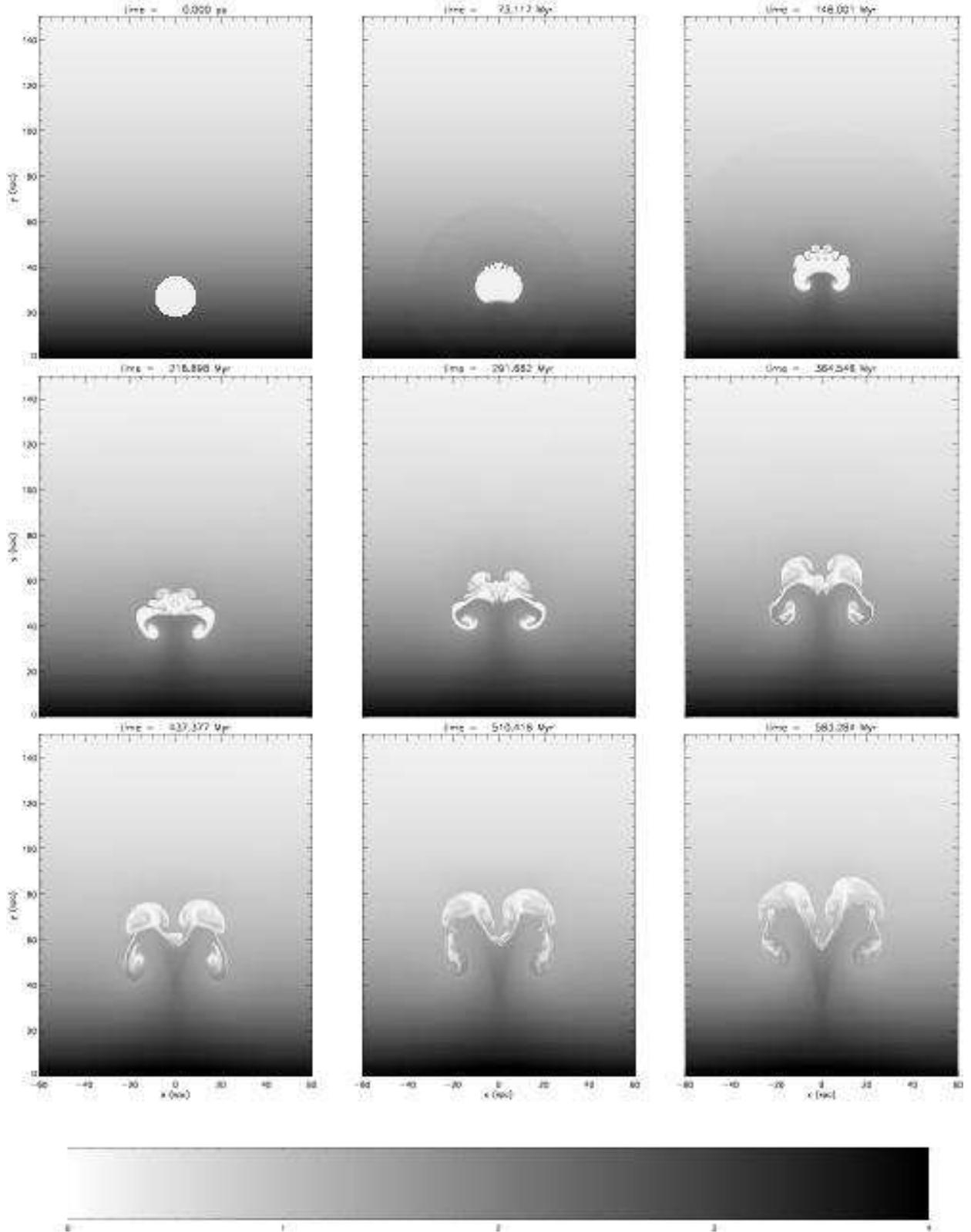}
\caption{The evolution of a bubble in a vertically magnetized background plasma, $\beta_0 = 462$.
The panels are plots of density.}
\label{fig:magy_high}
\end{figure}

\begin{figure}
\plotone{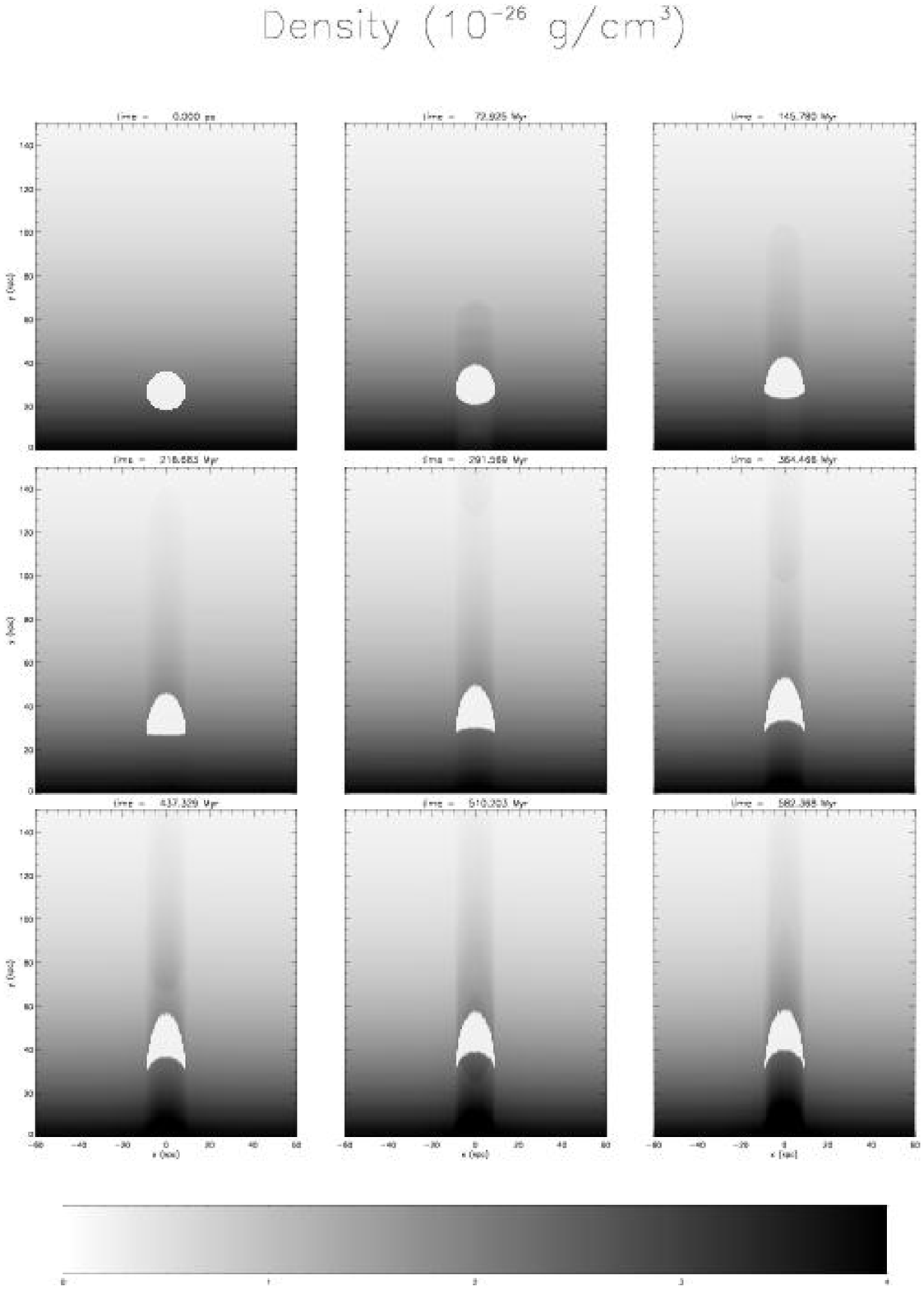}
\caption{The evolution of a bubble in a vertically magnetized background plasma, $\beta_0 = .019$.
The panels are plots of density.}
\label{fig:magy}
\end{figure}

A vertical field has a more direct dynamical effect on the rising bubble.
Adding a weak vertical field does not prevent the bubble from breaking
up, but somewhat suppresses the cascade of vortical motions at late
times by suppressing horizontal motions as compared to vertical motions
(Fig.~\ref{fig:magy_high}).  If we increase the strength of the magnetic
field, however, the suppression of horizontal motions can be so strong
that the bubble is deformed into a chevron and travels upwards intact
(Fig.~\ref{fig:magy}.)   A sufficiently strong vertical magnetic field
can completely contain the bubble, although it is difficult to imagine
a cluster having a magnetic field of this magnitude over the entire size
of the observed bubbles.

\section{MAGNETIC FIELDS: MAGNETIZED BUBBLE IN A BACKGROUND FIELD}
\label{sec:magneticbubbles}
In the previous sections, under-dense bubbles were supported by thermal
pressure.  This isotropic thermal pressure can support the bubble
against collapse, but not against being torn apart by the fluid motions
induced by its own rising.   The ambient magnetic fields could only
prevent this if the fields were strong enough (or oriented
fortuitously) to suppress those large-scale fluid motions.  We now
consider bubbles which are supported instead by magnetic fields, which
provide an anisotropic `pressure' support that also acts as a tension;
this tension can be enough to maintain the bubble even in the presence
of such motions.

One can construct families of solutions to the MHD equations with the
property that $\vec{B}$ is force-free (i.e.  $\vec{J} \times \vec{B} =
0$) everywhere in the domain except at the boundary between the
under-dense bubble and the ambient medium, where the Lorentz force
balances the gas pressure jump.  We choose a field used in
\cite{movingfluxtubes} for flux tubes:
\begin{eqnarray} 
B_\theta & = & \cases{B_b \frac{r}{a_0} & $r \le a_0$,\cr
                  0                 & $r > a_0$.\cr} \\
B_z  & = & \cases{B_b \sqrt{4 - 2 \left(\frac{r}{a_0}\right)^2} & $r \le a_0$,\cr
                   B_{z0}                                        & $r > a_0$.\cr} \\
B_b & = & \sqrt{\frac{8 \pi}{3} p_0 \left (1 - \frac{\rho_b}{\rho_0}\right) + B_{z0}^2}.  
\end{eqnarray}
Here $p_0$ is the ambient gas pressure, $\rho_0$ is the ambient gas
density, $\rho_b$ is the gas density inside the bubble, $B_{z0}$ is the
ambient magnetic field in the $\bf{\hat{z}}$ direction, and $a_0$ is
the radius of the bubble.  Thus we have a uniform magnetic field in the
${\bf{\hat{z}}}$ direction in the ambient medium, as discussed in our
previous section, but the bubble is also supported by an azimuthal
field and an enhanced ${\bf{\hat{z}}}-$field.

\begin{figure}[p]
\plotone{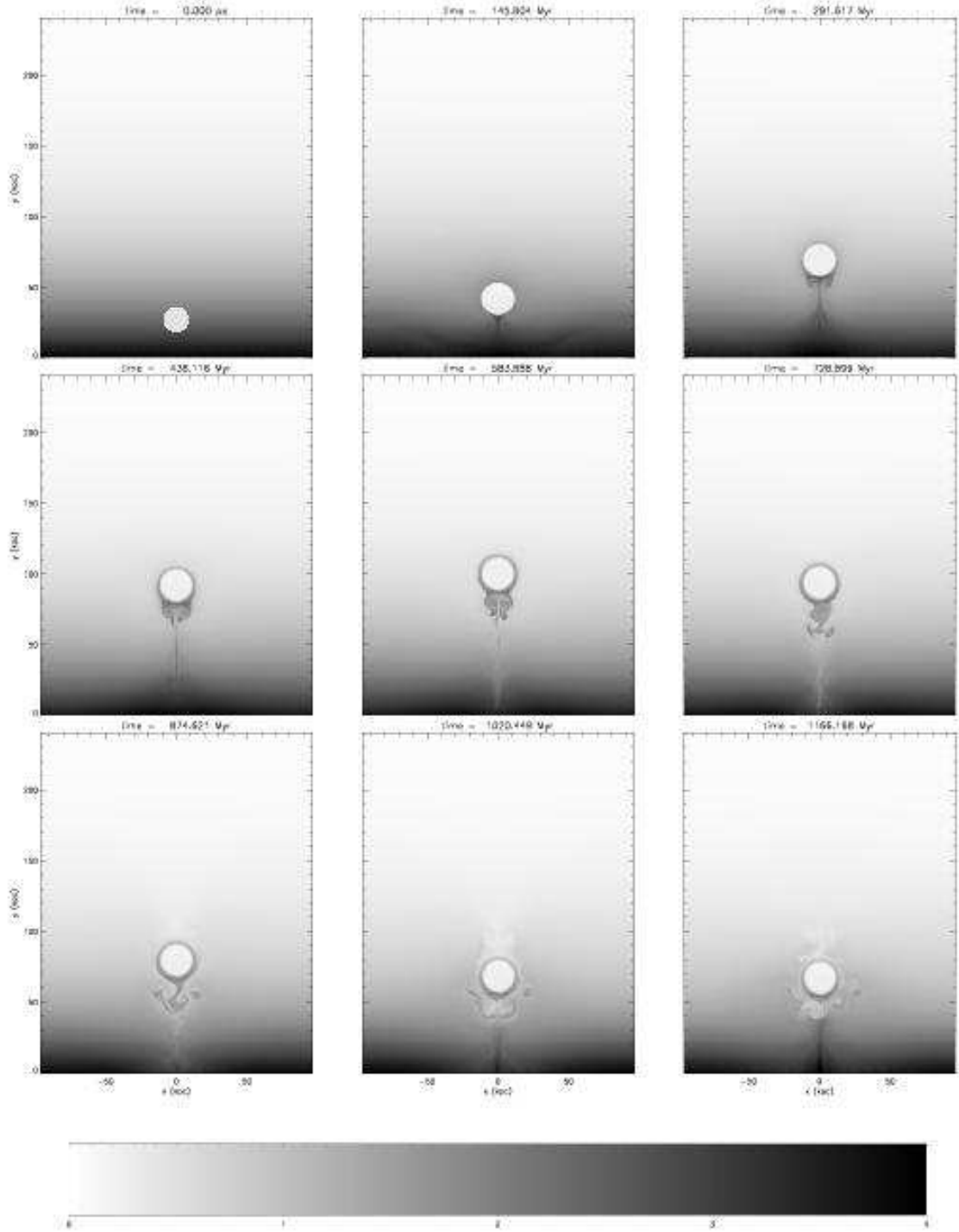}
\caption{The evolution of the magnetically supported bubble as it rises
without being torn apart.  The panels are plots of density.   This
simulation was run for a longer time; a larger domain was used so that
the bubble could rise without being interfered with by the top boundary.}
\label{fig:risingbubble}
\end{figure}

The hydrodynamic initial conditions are computed as in previous
sections, except that in this case the underdense region is at the same
temperature as the background material, and a constant Lorentz force at
the bubble interface supports the bubble against the larger external
ambient pressure.   The evolution is shown in
Fig.~\ref{fig:risingbubble}.  In this case, the bubble maintains its
form as it rises.

\begin{figure}[p]
\plotone{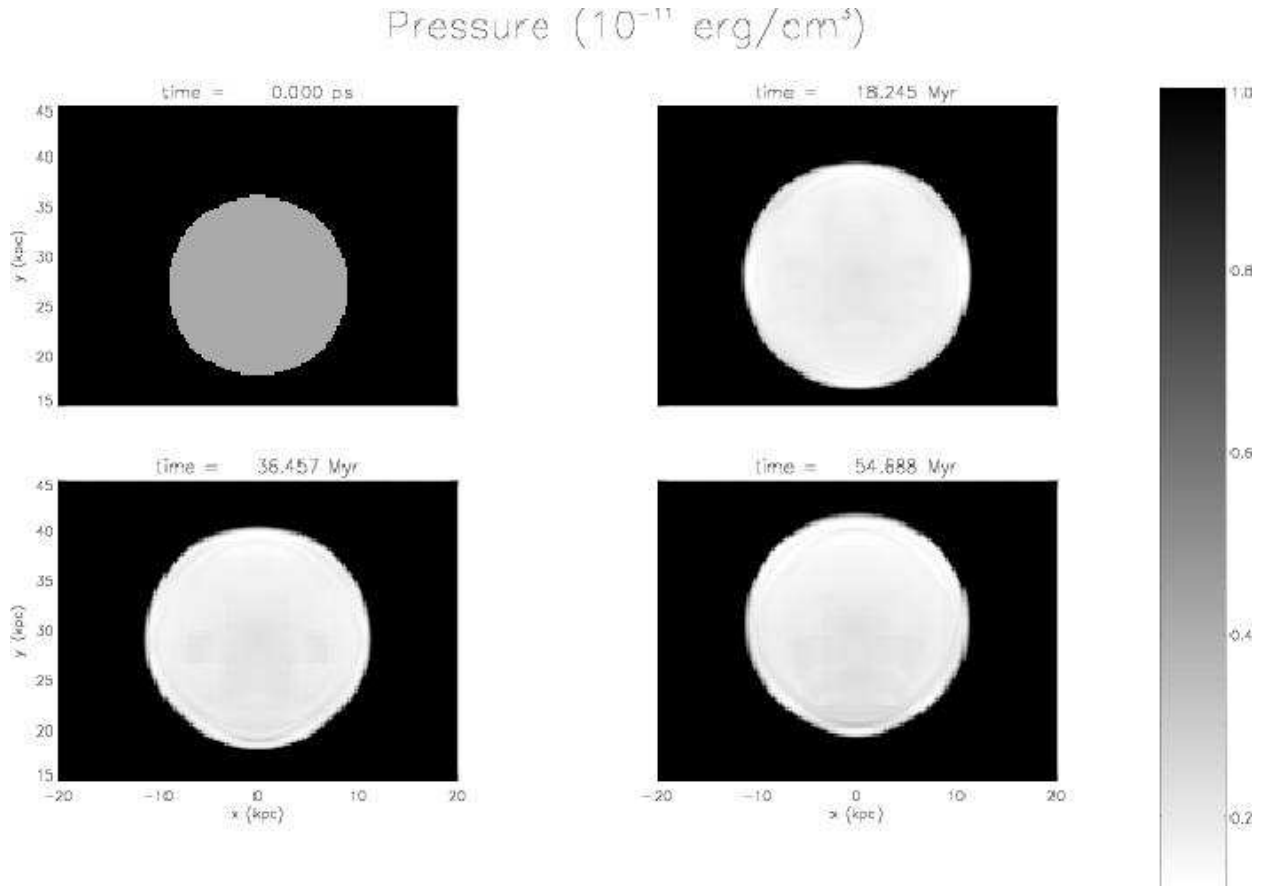}
\caption{The short-time evolution of the gas density inside the
magnetic bubble as it settles into an equilibrium.   The bubble 
expands and the gas inside the bubble is distributed towards the
bottom of the bubble.}
\label{fig:magbubblestruct}
\end{figure}

Because the ambient pressure changes by approximately 30\% over the
height of the bubble, while the bulk of the pressure support of the
bubble is fixed by the magnetic field jump, a transient occurs over the
course of a few sound-crossing times of the bubble (that is, a few times
44~\Myr) as the bubble equilibrates; the short-time evolution of gas
pressure is is shown in Fig.~\ref{fig:magbubblestruct}.  The magnetic
field is enough to maintain the bubble's shape until gas pressure and
density inside the bubble is redistributed to maintain equilibrium; we
see in Fig.~\ref{fig:magbubblestruct} that the bubble expands overall,
and gas slightly settles towards the bottom.

\begin{figure}
\plotone{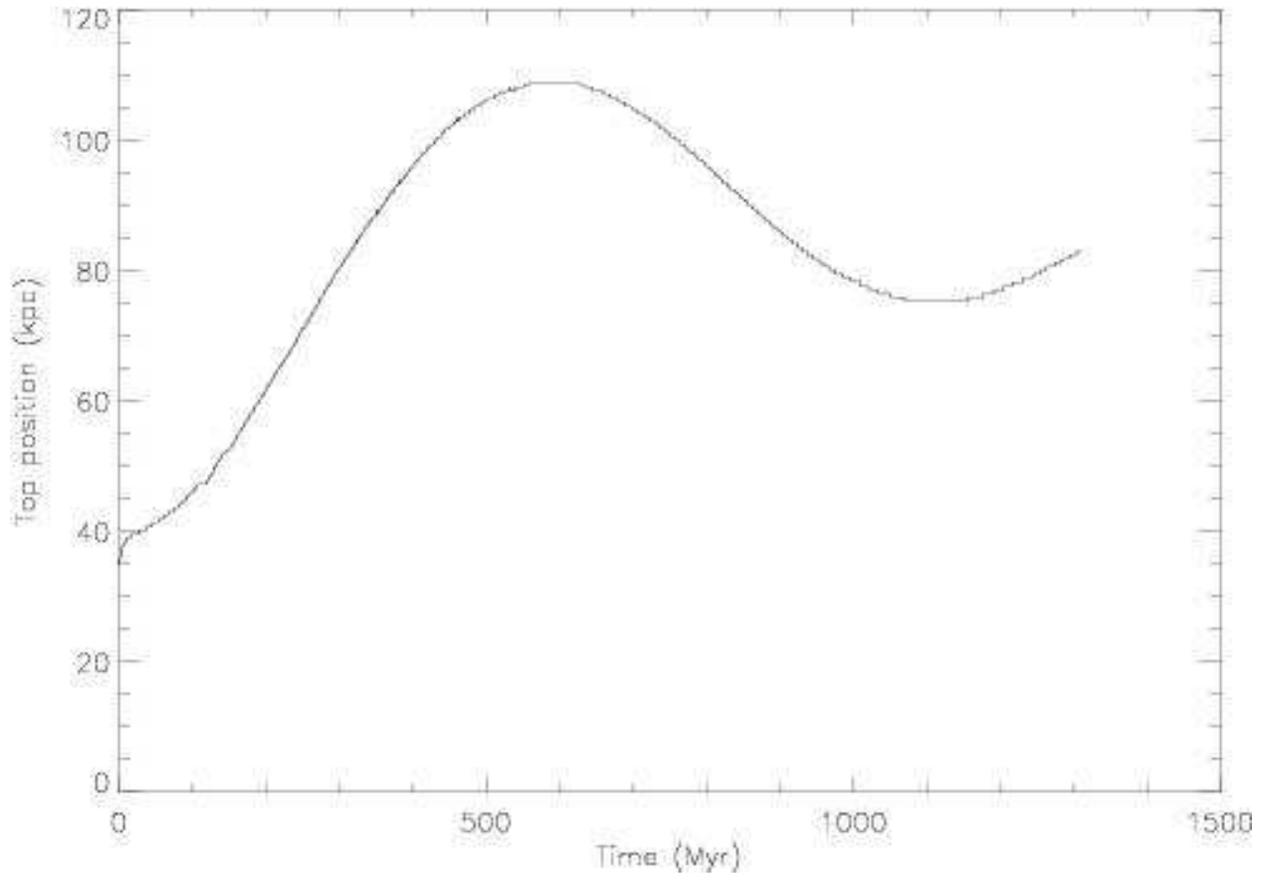}
\caption{The magnetically supported bubble rises until it passes it's point of neutral buoyancy then falls again, and again overshoots.}
\label{fig:risingbubblespeed}
\end{figure}

The bubble adiabatically expands as it rises, and eventually
overshoots the height of neutral buoyancy, decelerates, and falls
back again (Fig.~\ref{fig:risingbubblespeed}).  We also see again at
late times the development of an under-dense wake as in the uniform
${\bf{\hat{z}}}$-field case.

\begin{figure}
\plottwo{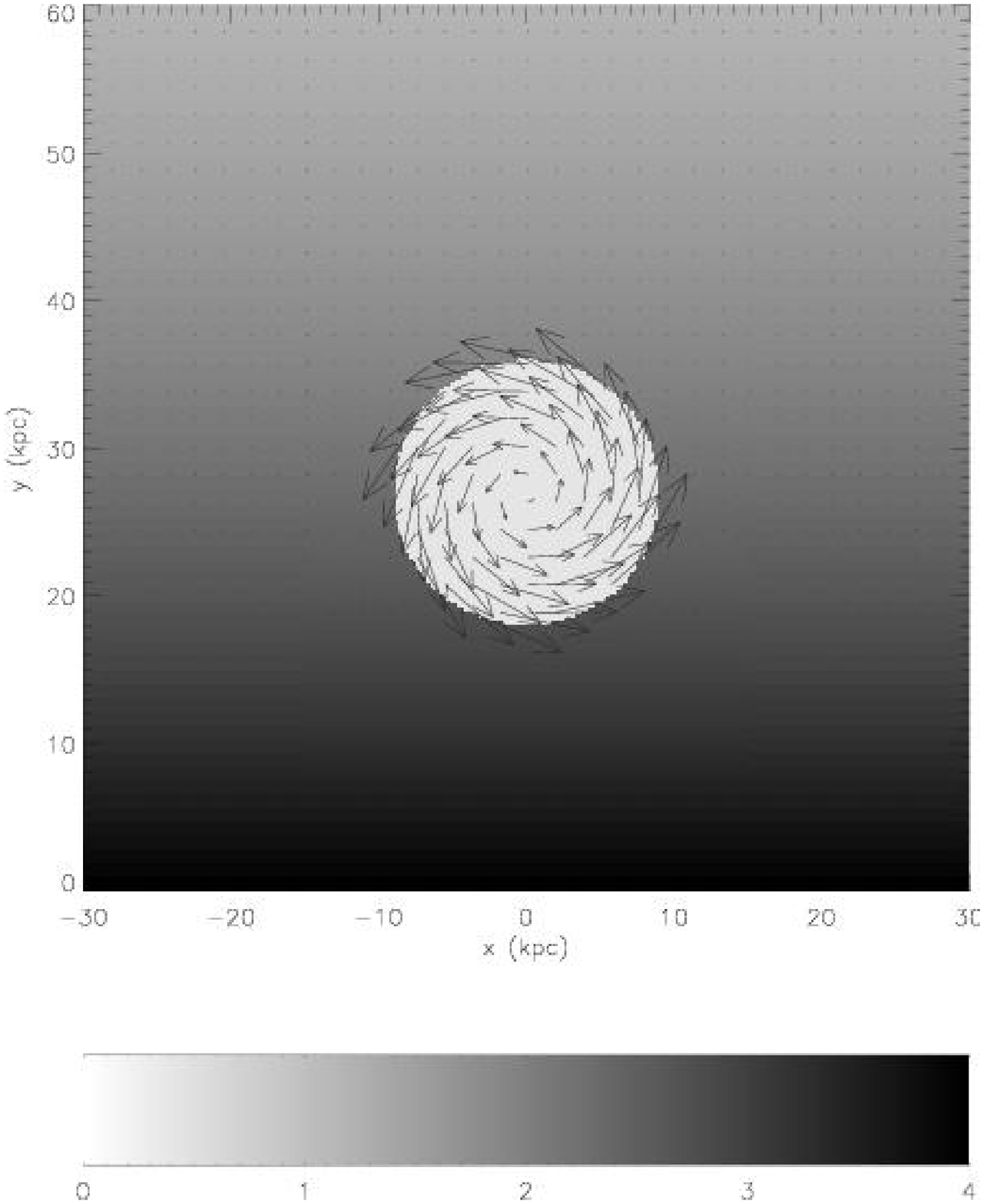}{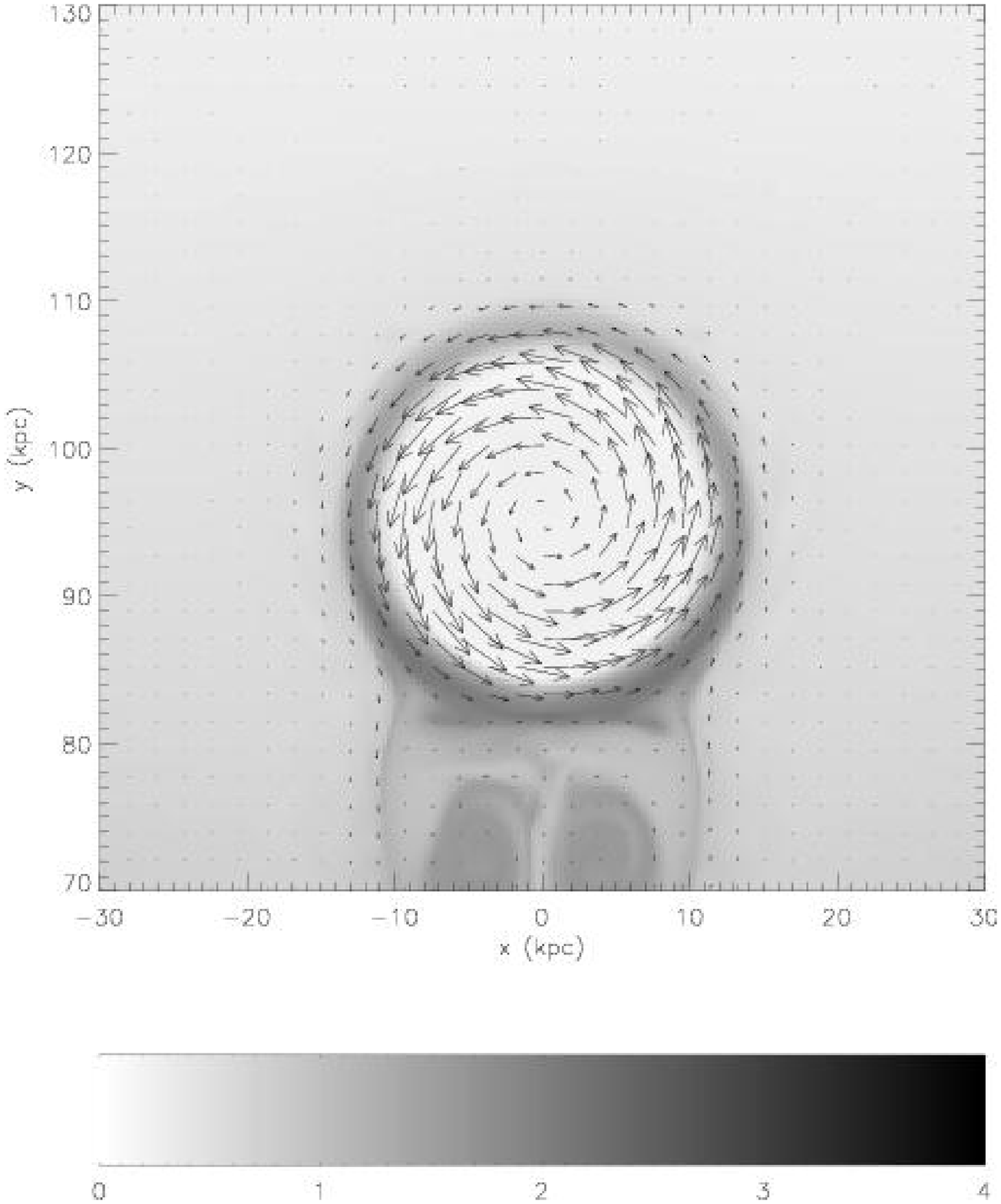}
\caption{Azimuthal field vectors, plotted over the density field, 
inside bubble and collar at t=0 and t=580 Myr.}
\label{fig:fields}
\end{figure}

\begin{figure}
\plotone{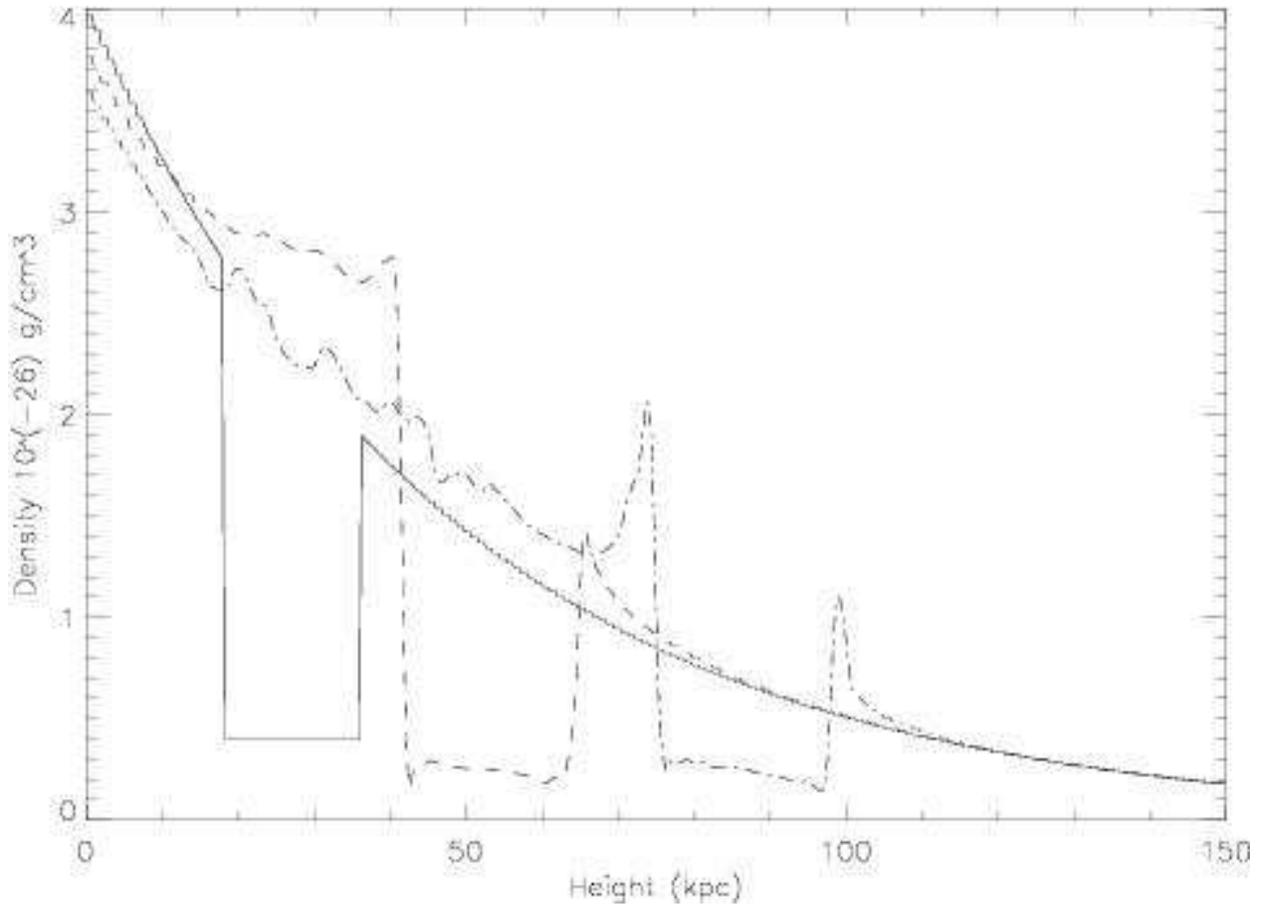}
\caption{Vertical density profiles (through $x = 0$) of the magnetized bubble simulation, taken at times $t = 0, 220, 440 \Myr$
         (solid line, dashed line, dot-dashed line).} 
\label{fig:mag-dens-profiles}
\end{figure}

Also notable is that the bubble carries a ring of higher density
material (Fig.~\ref{fig:fields}).  Such a high-density ring has also
been noted in Chandra observations of bubbles.  This collar of material
appears to be ``grabbed'' by the azimuthal field at time t=0 when the
entire bubble is surrounded by higher density material.  As shown in
Fig.~\ref{fig:mag-dens-profiles}, the high density collar surrounding
the low-density bubble does not change in density significantly as the
bubble moves upwards, suggesting that the bubble simply advects ambient
material.

\section{CONCLUSIONS}
\label{sec:conclusions}
This work focused on examining the primary physical conditions that
have to occur for the continued existence of the dark spots in galaxy
clusters found by the {\it Chandra} X-ray observatory.  We confirm that
bubbles without supporting magnetic fields are torn apart by
instabilities and vortical motions before they can move an entire
bubble height.   This is surprising, as `ghost' bubbles --- not radio
bright, and thus presumably no longer powered --- are observed
\citep{mazzotta,fabian2000,mcnamara2000} which are a significant
distance in bubble lengths away from the radio source which presumably
formed them.  The existence of these bubbles can be explained if they
are supported by an internal magnetic field.

Furthering this argument is the importance of thermal diffusion absent
any magnetic field effects.   One can numerically integrate the
diffusion equation using the thermal diffusivity quoted in
\S\ref{sec:initialmodels} given by \cite{spitzer}.   On doing so, one
finds that an $r = 9~\kpc$ bubble at a temperature of 100~\keV{} in an
ambient gas of 1~\keV{} is largely diffused away after 1~\Myr{}; a
10~\keV{} bubble survives for perhaps 150~\Myr.  Indeed, if the Spitzer
diffusivity is appropriate, there are constraints much tighter than
this; the Chandra observations show a sharp edge to the bubble, and
this edge would be blurred by thermal diffusion on time scales orders
of magnitude shorter than those required to completely dissipate the
bubble.

Thermal arguments alone, however, are not enough to require a magnetic
field to support the bubble; a weak tangled magnetic field may reduce
the conductivity without being strong enough to support the bubble, for
instance, or one could require more sophisticated diffusivities in the
presence of such large gradients than the Spitzer model.   In addition,
the synchrotron emission in the bubble may suggest that a 10-100~\keV{}
thermal gas is an insufficient model for the gas inside the bubble.

This research also suggests that the ring of brighter material
surrounding these bubbles may be caused by magnetic diffusion of the
field that maintains them.  There is also numerical evidence suggesting
a wake left behind the bubble as it moves.  Searching the radio
emissions for such a wake would be a good indicator as to whether or
not these bubbles are moving.

Since magnetic fields may be necessary to keep the bubbles intact as
they travel, future work should focus on MHD simulations, and in
particular performing simulations with more physically meaningful
geometry than 2D planar symmetry.   As was seen in the hydrodynamic
case, the difference between cylindrical and planar geometry
significantly altered the morphology, if not the timescale,  of the
disruption of the bubble; the difference in geometries would be even
more pronounced in the presence of magnetic fields, since the
orientation of the field also plays a role.   The hot bubble in
magnetic field simulations generalize easily to 3D; the magnetized
bubble simulation will be more complicated, as one can not write down
an analytical magnetic field in 3D which would support a spherical
bubble analogous to the cylindrical flux-tube presented here.   Such a
field, numerically generated, however, should also be able to support
the bubble without disruption.    Including the divergent geometry
appropriate to the center of a cluster will also be a necessary step.

\acknowledgements

Support for this work was provided by DOE grant number B341495 to the
ASCI/Alliances Center for Astrophysical Thermonuclear Flashes at the
University of Chicago.  K. Robinson was supported by the NSF REU
program at the University of Chicago.  LJD was supported by the
Department of Energy Computational Science Graduate Fellowship Program
of the Office of Scientific Computing and Office of Defense Programs in
the Department of Energy under contract DE-FG02-97ER25308.  MZ is
supported by the Scientific Discovery through Advanced Computing
(SciDAC) program of the DOE, grant number DE-FC02-01ER41176 to the
Supernova Science Center/UCSC.  All simulations were performed with
\FLASH{} 2.2.  The authors thank T.  Emonet for very useful discussions,
and the anonymous referee whose comments improved this work.
\FLASH{} is freely available at \url{http://flash.uchicago.edu/}.

\bibliographystyle{plainnat}
\bibliography{bubbles}

\clearpage

\clearpage

\clearpage

\clearpage

\clearpage

\clearpage

\begin{table}
\begin{center}
\begin{tabular}{rlll}
\tableline
\tableline
Time (Myr) & $v_s$ & $\frac{B_{s}}{B_0} $ & $\frac{B_{s}}{B_{\mathrm{sta ble}}} $ \\
\tableline
  73 &  0.014 &  1.1 &   $3.51\times 10^4$ \\
 150 &  0.027 &  2.8 &   $9.58\times 10^4$ \\
 220 &  0.130 &  3.8 &   $2.86\times 10^4$ \\
 290 &  0.109 &  6.4 &   $3.56\times 10^4$ \\
 370 &  0.090 &  6.1 &   $3.46\times 10^4$ \\
 440 &  0.094 &  5.7 &   $3.12\times 10^4$ \\
 510 &  0.081 &  3.0 &   $1.71\times 10^4$ \\
 580 &  0.066 &  2.0 &   $3.14\times 10^4$ \\
\tableline
\end{tabular}
\end{center}
\caption{Shearing and stability data for horizontal $\beta_0 = 462$ simulation.   Tabulated
is the time for which the data was taken, a mean shear velocity (in \kpc/\Myr), 
a mean enhancement of the magnetic field at the shear layer, and the factor by which
the magnetic field is greater than the stabilizing magnetic field.}
\label{tab:stabilize}
\end{table}

\newpage

\end{document}